\documentclass[10pt,journal,compsoc]{IEEEtran}

\usepackage{booktabs} % For formal tables
\usepackage{graphicx}
\usepackage{multirow}
\usepackage{amssymb,amsmath,amsfonts}
\usepackage{color}
\usepackage{booktabs}
\usepackage{hyperref}
\usepackage[ruled,linesnumbered,vlined]{algorithm2e}
\usepackage[switch]{lineno}
\usepackage{subcaption}
\usepackage{mathabx}
\usepackage{siunitx}
\newcommand{\nosemic}{\renewcommand{\@endalgocfline}{\relax}}% Drop semi-colon ;
\newcommand{\dosemic}{\renewcommand{\@endalgocfline}{\algocf@endline}}% Reinstate semi-colon ;
\let\oldnl\nl% Store \nl in \oldnl
\newcommand{\nonl}{\renewcommand{\nl}{\let\nl\oldnl}}% Remove line number for one line

\ifCLASSOPTIONcompsoc
  \usepackage[nocompress]{cite}
\else
  \usepackage{cite}
\fi

\newcommand{\modelname}{JTCR\xspace}

\begin{document}

\title{A Joint Two-Phase Time-Sensitive Regularized Collaborative Ranking Model for Point of Interest Recommendation}

\author{Mohammad~Aliannejadi, Dimitrios~Rafailidis, and Fabio~Crestani
\IEEEcompsocitemizethanks{
\IEEEcompsocthanksitem M. Aliannejadi and F. Crestani are with the Faculty of Informatics, Universit{\`a} della Svizzera italiana (USI), Lugano, Switzerland. \protect\\
E-mail: \{mohammad.alian.nejadi, fabio.crestani\}@usi.ch.
\IEEEcompsocthanksitem D. Rafailidis is with the Maastricht University, Maastricht, the Netherlands. \protect\\
E-mail: dimitrios.rafailidis@maastrichtuniversity.nl
}
}

\IEEEtitleabstractindextext{%hvi 
\begin{abstract}

The popularity of location-based social networks (LBSNs) has led to a tremendous amount of user check-in data. Recommending points of interest (POIs) plays a key role in satisfying users’ needs in LBSNs. While recent work has explored the idea of adopting collaborative ranking (CR) for recommendation, there have been few attempts to incorporate temporal information for POI recommendation using CR. In this article, we propose a two-phase CR algorithm that incorporates the geographical influence of POIs and is regularized based on the variance of POIs popularity and users’ activities over time. The time-sensitive regularizer penalizes user and POIs that have been more time-sensitive in the past, helping the model to account for their long-term behavioral patterns while learning from user-POI interactions. Moreover, in the first phase, it attempts to rank visited POIs higher than the unvisited ones, and at the same time, apply the geographical influence. In the second phase, our algorithm tries to rank users’ favorite POIs higher on the recommendation list. Both phases employ a collaborative learning strategy that enables the model to capture complex latent associations from two different perspectives. Experiments on real-world datasets show that our proposed time-sensitive collaborative ranking model beats state-of-the-art POI recommendation methods.

\end{abstract}

}

\IEEEpeerreviewmaketitle
\maketitle

\IEEEraisesectionheading{\section{Introduction}\label{sec-introduction}}

\IEEEPARstart{W}{ith} the availability of \textit{location-based social networks} (LBSNs), such as Yelp, TripAdvisor, and Foursquare, users can share check-in data using their mobile devices. LBSNs collect valuable information about users' mobility records with check-in data. Generating \textit{points-of-interest} (POIs) recommendations plays a crucial role in satisfying the user needs, for example when exploring a new POI or visiting a city~\cite{DBLP:journals/umuai/ChenCW15}. In fact, every town has numerous POIs and a user may have visited only a few in her hometown as well as when she is out-of-town~\cite{DBLP:conf/cikm/YuanCS14}.
POI recommendation tries to ensure user's satisfaction by suggesting her the most interesting locations in her vicinity, taking into account her preferences and contextual constraints \cite{DBLP:journals/tois/AdomaviciusSST05,DBLP:conf/sac/AliannejadiMC17}. 

Many challenges limit the accuracy of POI recommendation. For instance, a major challenge in POI recommendation is data sparsity~\cite{DBLP:journals/tkde/AdomaviciusT05,DBLP:journals/geoinformatica/0003ZWM15}.
Despite the fact that LBSNs feature a huge number of locations with a large variety, in practice users visit a very limited number of locations, making the user-item matrix sparse~\cite{DBLP:conf/www/ZhengZXY10}.  
Moreover, since users spend most of their time in their hometown~\cite{DBLP:conf/icde/LevandoskiSEM12}, the data sparsity problem is aggravated when a user visits a new city where she has no history of visited locations~\cite{DBLP:conf/cikm/FerenceYL13}. Several studies seek to address the data sparsity problem by incorporating additional information into the model, such as geographical and temporal information~\cite{DBLP:conf/recsys/GriesnerAN15,DBLP:conf/sigir/LiCLPK15,DBLP:conf/kdd/YeSLYJ11,DBLP:conf/airs/AliannejadiMC16}.
In fact, the data sparsity problem is even worse when recommending POIs as opposed to other items such as movies or songs. This mainly happens because the check-in data provides an implicit feedback~\cite{DBLP:conf/kdd/LianZXSCR14}, whereas users usually express their opinion about movies or songs with different ratings. \cite{DBLP:conf/sigir/LiCLPK15} argued that check-ins offer only positive examples whereas POIs with no check-in remain undiscovered which can potentially be interesting to the user. However, we argue that if a user has visited a place only once, we cannot simply infer a positive feedback but we can also infer what types of locations the user is interested in. In the case of places with higher check-in frequency, we assume that they are more preferred than those with fewer check-ins~\cite{DBLP:conf/sigir/LiCLPK15}.

In relevant literature, item recommendation is often treated as a rating prediction or matrix completion task~\cite{DBLP:conf/nips/SalakhutdinovM07}. However, \cite{DBLP:conf/www/Christakopoulou15} argued that considering the square loss as a measure of prediction effectiveness is not accurate. In other words, being able to produce a more accurate ranked list to a user should be rewarded. \textit{Collaborative Ranking} (CR) is based on this idea and focuses on the accuracy of recommendation at the top of the recommendation list for each user~\cite{DBLP:conf/wsdm/BalakrishnanC12}. Much work has adopted CR using explicit feedback from users such as item ratings.  For example, \cite{DBLP:conf/nips/WeimerKLS07} and \cite{DBLP:conf/www/Christakopoulou15} optimized a ranking loss to recommend movies to users. \cite{DBLP:conf/uai/RendleFGS09} showed that ranking-based learning of recommendation is also effective when one is dealing with implicit feedback from users. However, exploring CR for POI recommendation using implicit feedback is challenging because an effective strategy for sampling unvisited venues based on users' mobility and POIs proximity is required as part of the learning process.

Several studies started to incorporate temporal information to improve POI recommendation~\cite{DBLP:conf/sigir/YuanCMSM13,DBLP:conf/recsys/GaoTHL13,DBLP:conf/kdd/LiuLLQX16}. For example, temporal information has been incorporated in an hourly basis in \cite{DBLP:conf/sigir/YuanCMSM13} and in a sequential manner in \cite{DBLP:conf/ijcai/FengLZCCY15} to recommend the ``next" POI. 
Also, more advanced models considered both time and the sequential order of check-ins \cite{DBLP:conf/kdd/LiuLLQX16}. However, the long-term dynamics of users check-in behavior and venues popularity still need to be analyzed.
For example, an open-air bar is popular mainly during summertime, or a student is supposedly more active during her holidays. 

In this article, we propose a two-phase CR algorithm for POI recommendation. Our model is inspired by successful results of CR in other domains with explicit user feedback~\cite{DBLP:journals/jmlr/Rudin09, DBLP:conf/nips/WeimerKLS07} and of ranking methods using implicit user feedback for POI recommendation~\cite{DBLP:conf/sigir/LiCLPK15,DBLP:conf/uai/RendleFGS09}. 
We use a two-phase implicit feedback inference in our algorithm. In fact, we assume that single check-in means that a user ``likes" the POI and multiple visits mean that user prefers the POI more. Based on this assumption, in the first phase we push POIs with single or multiple check-ins at the top of the recommendation list, taking into account the geographical influence of POIs in the same neighborhood. In the second phase, we push POIs with multiple check-ins over the ones with a single visit. As argued in~\cite{DBLP:conf/sigir/LiCLPK15}, considering both visited and unvisited POIs in the learning, alleviates the sparsity problem. Therefore, the first phase of our algorithm addresses the sparsity problem, whereas the second stage boosts the accuracy of our model by pushing more relevant POIs at the top of the list. 
To take into account the dynamics of user and place check-in, we introduce a \textit{time-sensitive regularizer} in the ranking loss. The regularizer models the activity pattern of every user and venue over time. Adding this regularizing parameter to the objective function fuses the activity patterns into the ranking function in a collaborative way. 

In summary, our contribution in this article can be summarized as follows:
\begin{itemize}
    \item  We perform an extensive analysis to demonstrate the underlying patterns of preference and popularity over time.
    \item  We propose a general time-sensitive regularizer, taking into account the variance of users activities and venues popularity over time.
    \item  We propose a novel two-phase CR-based POI recommendation algorithm incorporating users implicit check-in feedback with a focus on the top of the list.
    \item  We propose a geographical similarity measure and add its influence to the model's objective function.
\end{itemize}

The experiments on two benchmark datasets show that the proposed approach outperforms state-of-the-art POI recommendation and CR methods.  In particular, we show that the joint learning strategy enables the model not only to address the sparsity problem but also to rank relevant POIs higher. The first phase mainly addresses the sparsity problem by adding the geographical influence as well as considering both visited and unvisited venues for training. The second phase improves the accuracy of the model, pushing to the top of the recommendation list the POIs that users prefer more. Moreover, we demonstrate the effectiveness of the time-sensitive regularizer, that is applied to both phases of the algorithm taking into consideration the long-term behavior of users and the popularity of POIs.

The remainder of the article is organized as follows, Section~\ref{sec-relatedwork} briefly reviews the related work while a deep analysis on the datasets is performed in Section~\ref{sec-analysis}. We describe our method in Section~\ref{sec-method} and in Section~\ref{sec-experiments}, evaluate the performance of our model against competitive models. Finally, Section~\ref{sec-conclusion} concludes the article.
\section{Related Work}
\label{sec-relatedwork}
The related work can be divided to the following topics: POI recommendation, collaborative ranking, and time-aware recommendation.

\subsection{POI Recommendation}

POI Recommendation plays an important role in satisfying users' expectations on LBSNs. Much work has been carried out in this area based on the core idea that users with similar behavioral history tend to act similarly~\cite{DBLP:journals/cacm/GoldbergNOT92}. This is the underlying idea of \textit{collaborative-filtering-based} (CF-based) approaches~\cite{DBLP:conf/recsys/GriesnerAN15,DBLP:journals/tois/YinCSHC14,DBLP:conf/cikm/FerenceYL13}. 
CF can be divided into two categories: memory-based and model-based. Memory-based approaches consider user rating as a similarity measure between users or items~\cite{DBLP:conf/www/SarwarKKR01}. Model-based approaches, on the other hand, employ techniques like matrix factorization~\cite{koren2008factorization}.
However, CF-based approaches often suffer from data sparsity since there are a lot of available locations, and a single user can visit only a few of them. As a consequence, the user-item matrix of CF becomes very sparse, leading to poor performance in cases where there is no significant association between users and items. 
Many studies have tried to address the data sparsity problem of CF by incorporating additional information into the model~\cite{DBLP:conf/sigir/YeYLL11,DBLP:conf/sigir/YuanCMSM13}. More specifically, \cite{DBLP:conf/sigir/YeYLL11} argued that users check-in behavior is affected by the spatial influence of locations and proposed a unified location recommender system incorporating spatial and social influence to address the data sparsity problem. \cite{DBLP:journals/tois/YinCSHC14} proposed a model that captures user interests as well as local preferences to recommend locations or events to users when they are visiting a new city. 
\cite{DBLP:conf/cikm/YuanCS14} proposed to consider both geographical and temporal influences while recommending POIs to the users via a geographical-temporal influences aware graph. They proposed to propagate these influences using a breadth-first strategy.
\cite{DBLP:conf/cikm/FerenceYL13} took into consideration user preference, geographical proximity, and social influences for POI recommendation. \cite{DBLP:journals/tist/ChengYKL16} proposed a multi-center Gaussian model to capture users' movement pattern as they assumed users' movements consist of several centers. In a more recent work, \cite{DBLP:journals/tois/ZhangLW16} considered three travel-related constraints (i.e., uncertain traveling time, diversity of the venues, and venue availability) and use them to prune the search space. \cite{DBLP:conf/recsys/GriesnerAN15} also proposed an approach integrating temporal and geographic influences into matrix factorization. \cite{DBLP:conf/ecir/Aliannejadi17} proposed a probabilistic mapping approach to determine the most salient information from a venue's content to reduce dimensionality of data.  More recently, \cite{alianSigir17} considered the appropriateness of a POI given a user's context while ranking the venues. 
\cite{DBLP:journals/tist/ZhangDCLZ13} fused virtual ratings derived from online reviews into CF.  
\cite{DBLP:journals/tist/CuiSNHM17}  investigated how geotagged photos can be linked to venues to study users' tastes. Finally, \cite{DBLP:conf/ictai/YuanJGCYA16} addressed the data sparsity problem assuming that users tend to rank higher the POIs that are geographically closer to the one that they have already visited. 
Differently from these studies, our work addresses the data sparsity problem by including the unvisited venues in the learning process. We consider unvisited venues as negative training examples based on the assumption that a user ``likes" and prefers venues that she has visited over the ones that she has not~\cite{DBLP:conf/kdd/LiuLLQX16}.

\subsection{Collaborative Ranking}

Another line of research lies in combining the ideas of CF and \textit{Learning to Rank} (LTR).  
LTR methods have been proved to be effective in \textit{Information Retrieval} (IR)~\cite{DBLP:journals/ftir/Liu09}. LTR learns a ranking function which can predict a relevance score given a query and document. There are three categories of LTR, namely, point-wise~\cite{DBLP:conf/nips/CrammerS01}, pair-wise~\cite{rankNet}, and list-wise~\cite{listNet}. In short, point-wise approaches predict ranking scores for individual items. Pair-wise approaches, on the other hand, learn the order of the items. List-wise approaches consider an entire list of items as individual training example. CR takes the idea of predicting preference order of items from LTR and combines it with the idea of learning the loss function in a collaborative way. \cite{DBLP:conf/nips/WeimerKLS07} used a surrogate convex upper bound of \textit{Normalized Discounted Cumulative Gain} (nDCG) error together with matrix factorization as the basic rating predictor. \cite{DBLP:conf/cikm/ShiKBLH13} explored optimizing a surrogate lower bound for \textit{Expected Reciprocal Rank} (ERR) for data with multiple levels of relevance. \cite{DBLP:conf/www/Christakopoulou15} followed the idea of pair-wise LTR approaches. In particular, the authors base their work on LTR methods with an emphasis on the top of the recommendation list. This approach, however, is limited to explicit user feedback such as user ratings for movies. \cite{DBLP:conf/sigir/RafailidisC16a} presented an LTR model, taking into account the relevant items of users and their friends, pushing these items at the top of the list. \cite{DBLP:conf/uai/RendleFGS09} presented a generic optimization criterion as well as a learning algorithm for incorporating implicit feedback while learning personal ranking for users and they demonstrated its effectiveness on approaches such as matrix factorization. In~\cite{DBLP:conf/cikm/RafailidisC16}, authors combined various LTR methods into a joint model aiming to enhance the recommendation accuracy with trust relationships. In a more recent work, \cite{DBLP:conf/recsys/RafailidisC17} proposed a model considering not only relevant items of the user and her trusted friends, but also the items of her distrusted foes. Unlike the above works, our method uses a two-phase strategy to use the implicit feedback, i.e. check-ins. Also, we regularize the ranking loss based on users activity and venues popularity variances on a monthly basis. The social influence on POI recommendation, however, is left for future work.

\subsection{Time-Aware Recommendation}
Many researchers have studied the temporal influence on users' preferences. A group of studies conducts time-aware recommendation learning of users' temporal preference for specific time slots and for recommending POIs for a given time slot, like an hour of a day~\cite{DBLP:conf/cikm/DingL05}.  
\cite{DBLP:conf/sigir/YuanCMSM13} computed the similarity between users by finding the same POIs at the same time slots in their check-in history to train a user-based CF model. \cite{DBLP:conf/icdm/YaoFLLX16} matched the temporal regularity of users with the popularity of POIs to improve a factorization-based algorithm. \cite{DBLP:journals/tois/LiJHL17} proposed a time-aware personalized model adopting a fourth-order tensor factorization-based ranking which enables the model to capture short-term and long-term preferences.
\cite{DBLP:journals/tois/YinCZWHS16} proposed topic-region model that discovers the semantic, temporal, and spatial patterns of users' check-ins and uses the additional information to address the data-sparsity problem. 
\cite{DBLP:conf/sigmod/YinCCHH14} defined the temporal context as the public's attention at a certain time and proposed a temporal context-aware mixture model, modeling the topics related to users' interests and temporal context in a unified way. This work was later extended in \cite{DBLP:journals/tois/YinCCHZ15} to a dynamic temporal context-aware mixture model, capturing users’ evolving interests.
\cite{DBLP:conf/recsys/GaoTHL13} preserved the similarity of personal preference in consecutive time slots by considering different latent variables at each time slot for each user. 
However, the long-term behavior of users still needs to be studied.
For example, a high school student visits more places during summer, or a beach bar could be even shut down during winter. Therefore, it is crucial to consider the time-dependent activity patterns of both users and venues while training a recommender system. 
There also exists another category of approaches which tries to recommend the next POI to visit, also known as successive POI recommendation. For example, \cite{DBLP:conf/ijcai/ChengYLK13} captured sequential check-in behavior of users by training personalized Markov chain. \cite{DBLP:conf/kdd/LiuLLQX16} combined the ideas of both categories by recommending POIs for a particular time, exploiting sequential patterns of users. Our work distinguishes itself from these studies by considering a long-term activity shift of both users and locations in the course of time.
\section{Data Analysis}\label{sec-analysis}
In this section, we conduct an extensive analysis of two real-world datasets to explore user preferences' dynamics over time.

\subsection{Data}
We use two real-world check-in datasets from Foursquare and Gowalla provided by~\cite{DBLP:conf/sigir/YuanCMSM13}\footnote{\url{http://www.ntu.edu.sg/home/gaocong/data/poidata.zip}}. The Foursquare's dataset originally consists of 342,850 check-ins of users in Singapore in the period of Aug. 2010 and Jul. 2011~\cite{DBLP:conf/sigir/YuanCMSM13}. The Gowalla's dataset, on the other hand, includes 736,148 check-ins made by users in California and Nevada in the period of Feb. 2009 and Oct. 2010~\cite{DBLP:conf/sigir/YuanCMSM13}. Every check-in contains a user id, POI id, time, and geographical coordinates. For a fair comparison, we also use the preprocessed data as in~\cite{DBLP:conf/sigir/YuanCMSM13}. Users who have less than 5 check-ins, as well as POIs with less than 5 check-ins are removed from the datasets. Finally, we have 194,108 check-ins on Foursquare's and 456,988 check-ins on Gowalla's. We used the geographical coordinates of POIs to retrieve their corresponding categories such as bar, pizza place. More details are listed in Table~\ref{tab-stats}. Notice that multiple check-ins refer to the check-ins that were made by a particular user to a particular POI more than once. As shown in Table~\ref{tab-stats}, 45.51\% of check-ins on the Foursquare's dataset refer to users visiting POIs more than once. In Gowalla's, we observe fewer multiple check-ins, namely, 32.69\% of all check-ins. 

\begin{table}[t]
    \centering
    \caption{General statistics of the datasets}
    \setlength{\extrarowheight}{.4em}
    \begin{tabular}{l|l|l}
        \hline
         & \textbf{Foursquare's} & \textbf{Gowalla's} \\
        \hline
        \# of users & 2,231 & 10,162 \\\hline
        \# of POIs & 5,596 & 24,250 \\\hline
        \# of check-ins & 194,108 & 456,988 \\\hline
        Avg. POIs per user & 45.57 & 30.356 \\\hline
        Avg. users per POI & 18.90 & 12.69 \\\hline
        Multiple check-ins & 45.51\% & 32.69\% \\ \hline
        Density & 0.0081 & 0.0012 \\
        \hline
    \end{tabular}
    \label{tab-stats}
\end{table}

\subsection{Time-Dependency of User Activities and Interests}\label{sec-time-analysis}

\begin{figure}[t]
 \centering
	\begin{subfigure}[b]{\columnwidth}
        \includegraphics[width=\textwidth]{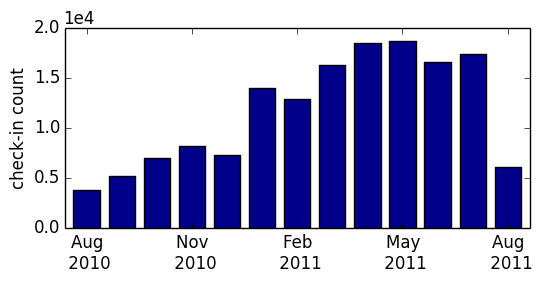}
        \caption{Foursquare's}
        \label{fig-hist-fsq}
    \end{subfigure}
	\begin{subfigure}[b]{\columnwidth}
        \includegraphics[width=\textwidth]{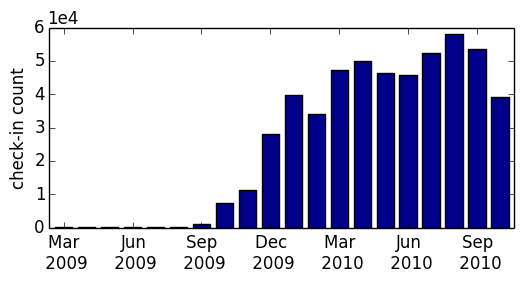}
        \caption{Gowalla's}
        \label{ffig-hist-gwl}
    \end{subfigure}
\caption{Number of check-ins per month on Foursquare's and Gowalla's.} 
\label{fig-hist}
\end{figure}

\begin{figure}[t] \centering
\includegraphics[width=\columnwidth]{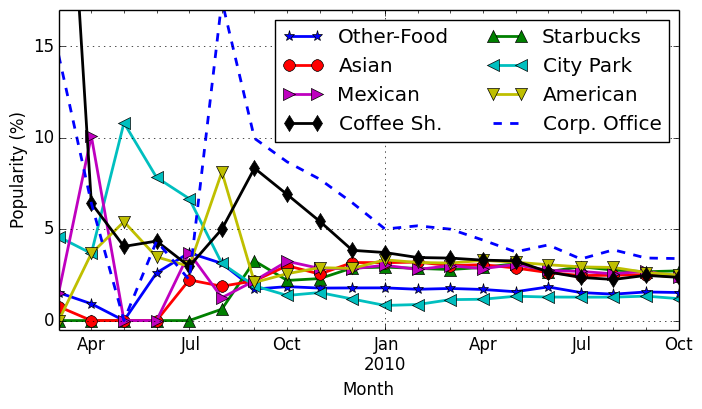}
\caption{Popularity of the top-8 categories over time on Gowalla's (best viewed in color).} 
\label{fig-popularity}
\end{figure}

Many studies analyzed users activity patterns over time~\cite{DBLP:conf/cikm/YuanCS14,DBLP:conf/sigir/YuanCMSM13,DBLP:conf/icdm/YaoFLLX16,DBLP:conf/recsys/GaoTHL13}. However, user activities have been mostly carried out at an hourly basis. We conduct a long-term user activity analysis, namely we study the user behavior over several months. This monthly analysis allows us to realize the long-term shift of users interest. Users' interests could evolve over time while sustaining their personal patterns. For instance, a teenager who reaches the legal age of drinking often starts going to bars (long-term interest shift), whereas an adult user often chooses to travel on Easter (personal pattern). \autoref{fig-hist} depicts the distribution of check-ins of both datasets every month. There is a significant shift of activity on each month. For instance, we observe that the number of check-ins increases from Feb. to Jul. in both datasets. 
Next, we report further analysis only on Gowalla's, as we observed similar attributes when analyzing the Foursquare's dataset. 
\autoref{fig-popularity} shows the popularity of the top-8 venue categories. As we can see, before Oct. 2009 we observe dramatic shift of popularity. However, after Oct. 2009, the top categories exhibit a rather more stable popularity pattern as the dataset grows. Interestingly, we still can observe popularity shifts between ``Coffee Shop", ``American", ``Mexican", ``City Park", and ``Asian". This suggests that users also exhibit a time-dependent pattern in visiting popular categories of POIs. 

\begin{figure*}[ht]
 \centering
	\begin{subfigure}[b]{0.49\textwidth}
        \includegraphics[width=\textwidth]{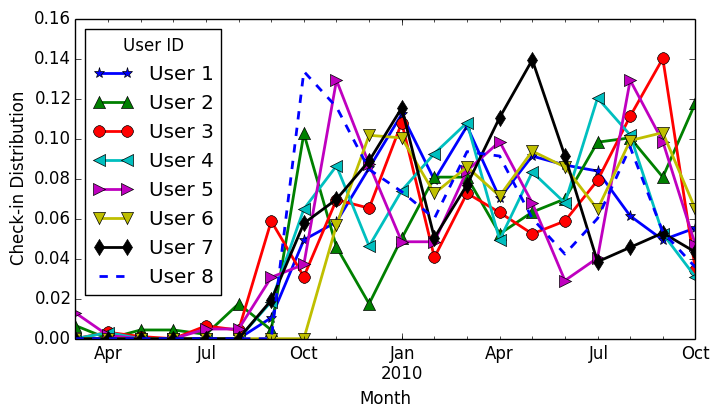}
        \caption{Least variant users}
        \label{fig-vars-fsq-least-u}
    \end{subfigure}
	\begin{subfigure}[b]{0.49\textwidth}
        \includegraphics[width=\textwidth]{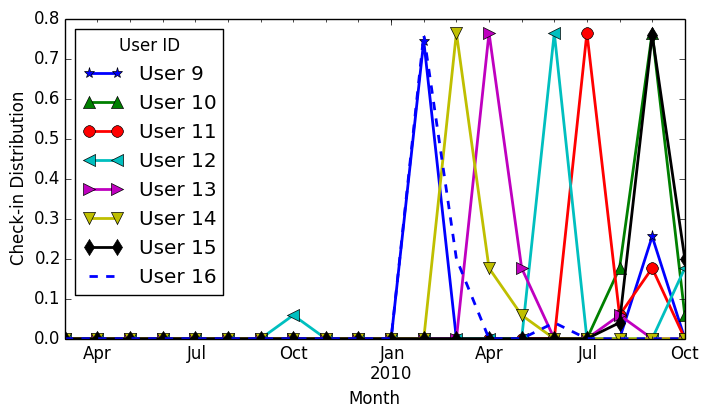}
        \caption{Most variant users}
        \label{fig-vars-most-u}
    \end{subfigure}
    \begin{subfigure}[b]{0.49\textwidth}
        \includegraphics[width=\textwidth]{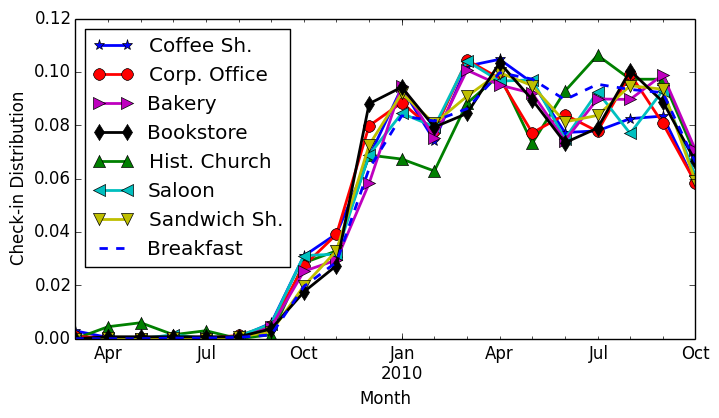}
        \caption{Least variant POI categories}
        \label{fig-vars-least-v}
    \end{subfigure}
	\begin{subfigure}[b]{0.49\textwidth}
        \includegraphics[width=\textwidth]{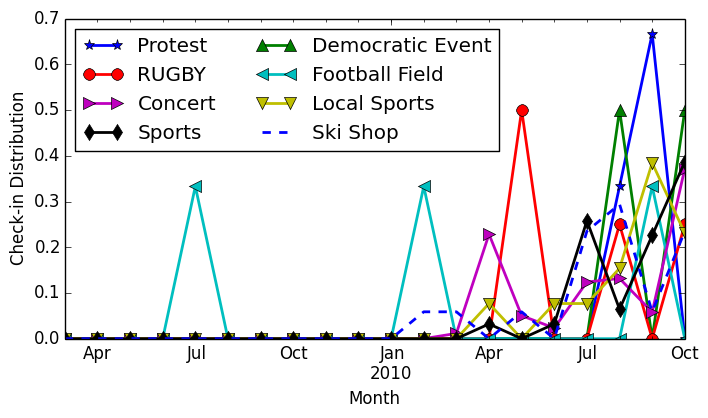}
        \caption{Most variant POI categories}
        \label{fig-vars-most-v}
    \end{subfigure}
\caption{Check-in distribution of users and POI categories over time in Gowalla's\textsl{•}. Figures a \& c depict the least time-variant users and POI categories, respectively. Figures b \& d, in contrast, show the distribution of the most time-variant users and POI categories, respectively (best viewed in color).} 
\label{fig-vars}
\end{figure*}

\autoref{fig-vars} shows the activity of two samples of users in a long period. It also shows how the popularity of two samples of POI categories change over time. Note that, all the plots of \autoref{fig-vars} illustrate the distribution of each user or category over time, that is, for each user or category, summing all the values of Y over time (X) equals to 1. We compute the variance of check-ins for both users and categories. Then, we pick the ones with the lowest in \autoref{fig-vars}a \& c and the ones with the highest variances in \autoref{fig-vars}b \& d. Our aim is to pick some representative user and categories from either end of the spectrum. As we can see, even users who are supposed to be the most stable users (\autoref{fig-vars}a) exhibit very time-dependent check-in patterns. It is worth noting that we ignored users with less than 50 check-ins in plotting \autoref{fig-vars}b, as those users mostly appear once in the dataset and hence are listed among the top variant users. However, we still observe that the users with high check-in variance are those who appear a few times and have less regular check-in behavior. This suggests that the more active a user is, the less variant her check-in pattern is. We also observed \textit{a negative correlation between the users' check-in variance and quantity} (Spearman: $-$0.5583) which supports our assumption. As for the POI categories, \autoref{fig-vars}c depicts the least variant categories. First, all of the categories relate to users' daily activities, for example going to a coffee shop or sandwich store. These categories are more time-independent since they are less affected by seasonal changes or weather conditions. Similar to what we observed for users, more popular categories are also less time-variant. In fact, we found \textit{a negative correlation between POI categories variance and popularity} (Spearman: $-$0.8411). On the other hand, the more time-variant categories (\autoref{fig-vars}d) are less popular and they depend mainly on weather and political events. For example, ``Ski Shop" only appears during winter and ``Democratic Event" appears as a political campaign takes place. 

All our observations suggest that there is an underlying temporal pattern in check-in activities from the perspective of both users and place categories. As we argued, check-in variance reveals meaningful information and thus we design our model focusing on these aspects.

\subsection{Users' Multiple Check-ins}\label{sec-multi}

\begin{figure*}[t]
    \centering
    \includegraphics[width=\textwidth]{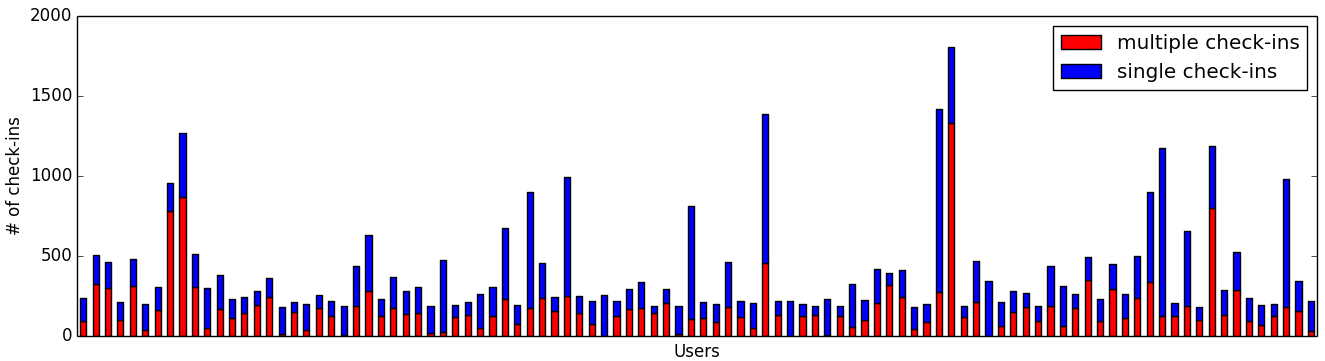}
    \caption{Check-in histogram of 100 randomly-sampled users from the 500 most active users of Gowalla's. For each user, the red bar denotes the number of multiple check-ins, while the blue bar denotes the number of single check-ins.} 
    \label{fig-multi}
\end{figure*}

As we discussed earlier in Section~\ref{sec-introduction}, a good indicator of a user's preference is the fact that she visits the same POI multiple times. 
A multiple check-in between a user $\rho$ and a venue $l$ occurs if $\rho$ visits $l$ more than once.
Other studies assumed multiple check-ins as an implicit feedback and demonstrated its effectiveness~\cite{DBLP:conf/sigir/LiCLPK15}. We also observed that a considerable amount of check-ins are multiple check-ins (\autoref{tab-stats}). We randomly picked 100 users from the 500 most active users of Gowalla's. \autoref{fig-multi} shows the quantity of single check-ins as well as multiple check-ins of the sample users. As we can see, users spend a considerable amount of their time visiting POIs they have visited before. In fact, multiple check-ins constitute 20\% of check-ins for an average user on Gowalla's (with a standard deviation of 21). According to \autoref{fig-multi}, the more active a user is, the more multiple check-ins she has. We calculated the correlation between the number of check-ins and the percentage of multiple check-ins per user. We observe that there is a positive correlation (Spearman: +0.4257) between the two variables, supporting our assumption. Based on these observations, we confirm that multiple check-ins provide valuable information regarding users' preference and since these account for a considerable amount of the datasets, we design our model with a focus on them.

\subsection{Remarks} \label{sec-intuitions}
Here, we present the main findings of our analysis. As observed in Section~\ref{sec-time-analysis}, users exhibit a long-term shift in their check-in behavior. As a consequence, POIs witness a shift in their popularity. However, many of these shifts are natural because users grow old and their habits evolve. Also, even though some types of POIs exhibit less time-sensitivity, many POIs are highly dependent on temporal phenomena such as seasonal changes or political campaigns. In addition, over a long period of time, less active users exhibit more variance in their check-in behavior. The same behavior is also observed for POIs, that is, less popular POIs suffer from more variance in terms of users' check-ins. Thus, it is crucial to take into consideration how variant a user or POI has been in the past while learning the model. 

According to \ref{sec-multi}, there is a positive correlation between the activity of a user and the number multiple check-ins she has had. This also means that a user who has visited the same POI several times in the past is more likely to visit the same POI in the future. Moreover, single check-ins are already crucial and should be considered in the model to account for users' preferences as they choose to visit a particular POI when they could have chosen other venues in the same neighborhood. This gives us a rough estimate of their interest and preferences while multiple check-ins are more accurate indicators of users' interest. Thus, both single and multiple check-ins need to be involved while learning the model.    

%%%%%%%%%%%%%%%%%%%%%%%%%%%%%%%%%%%%%%%%%%%%%%%%%%%%%%%%%%%%%%%%%%%%%%%%%%%%
\section{Proposed Method}\label{sec-method}
Following the remarks based on our analysis in Section \ref{sec-analysis}, in this section, we present our model.
We first introduce the notations of our model.
Let $\mathcal{P} = \{\rho_1, \dots, \rho_n\}$ be the set of $n$ users and $\mathcal{L} = \{l_1, \dots, l_m\}$ be the set of $m$ POIs. As an implicit feedback, we consider POIs that each user $\rho_i$ has visited, $\mathcal{L}_i$, as ``relevant" and all other unvisited POIs as ``irrelevant" in the neighborhood where the user has visited all relevant items. Also, we consider POIs with multiple check-ins as ``more relevant". Therefore, we define $\mathcal{L}^+_i$ as the set of relevant POIs, $\mathcal{L}^-_i$ as the set of irrelevant POIs, and $\mathcal{L}^*_i$ the set of ``more-relevant" POIs for user $\rho_i$. We also define $n^+ = |\mathcal{L}^+_i|$, $n^-=|\mathcal{L}^-_i|$, and $n^*_i=|\mathcal{L}^*_i|$.

Our aim is to define a ranking function $f_i(l)$ for each user $\rho_i$ to rank more-relevant POIs higher than relevant POIs, and relevant POIs higher than irrelevant POIs. Moreover, we intend to incorporate the influence of POIs that are located close to each other.
Let $U \in \mathbb{R}^{d \times n}$ be the latent factor for users and $V \in \mathbb{R}^{d \times m}$ be the latent factor for POIs. It is worth noting that $\mathbf{u}_i$ corresponds to $\rho_i$ and $\mathbf{v}_j$ corresponds to $l_j$. The ranking function for the $i^{\text{th}}$ user $\rho_i$ and the $j^{\text{th}}$ POI $l_j$ would be $f_i(l_j) = \mathbf{u}_i^T\mathbf{v}_j$.

We design a two-phase objective function, where the first phase constructs ranking functions $f_i(l)$ such that relevant POIs are ranked higher than irrelevant POIs. In the second phase, functions $f_i(l)$ are updated to rank more-relevant POIs higher than relevant POIs. In Section \ref{sec-geo}, we first describe how we compute the distance between two POIs and in Section \ref{sec-phase1} we explain the first phase of the objective function in which the geographical influence is also incorporated. In Section \ref{sec-phase2}, we give details on the second phase of the objective function. In Section \ref{sec-reg}, we describe the proposed time-sensitive regularizer which takes into consideration the users' long-term behavioral patterns, and in Section \ref{sec-opt} we present an overview of the proposed algorithm and how both phases of the objective function are optimized jointly.

\subsection{Geographical Similarity} \label{sec-geo}
 
We compute the geographical similarity between two POI to incorporate the geographical context while characterizing the user's geographical preferences. The similarity is inversely proportional to the distance between two POIs.
Inspired by the relevant studies~\cite{DBLP:conf/sigir/LiCLPK15,DBLP:conf/kdd/LianZXSCR14,DBLP:conf/iir/AliannejadiC18,DBLP:conf/ictir/AliannejadiRC18} where a simple geographical measure improved the models significantly, 
we use the Haversine formula to compute the angular distance between $l_i$ and $l_j$:

\begin{align*}
    \begin{split}
          & h_{ij} = 2 \times \\
          & \arcsin\Big(\sqrt{\sin^2(\Delta\phi_{ij}/2) + \cos\phi_i \times \cos\phi_j \times \sin^2(\Delta\eta_{ij}/2)}\Big)~,
    \end{split}    
\end{align*}
\noindent where $\phi_i$ and $\phi_j$ are latitudes of $l_i$ and $l_j$ in radian, respectively. Accordingly, $\eta_i$ and $\eta_j$ are longitudes of $l_i$ and $l_j$ in radian. Then we calculate the geographical similarity between $l_i$ and $l_j$ as follows:
\[
    g_{ij} = \frac{1}{1+ (h_{ij} \times R)}~,
\]

\noindent where $R$ is the earth's radius ($R=$6,371KM). In the following section, we demonstrate how $d_{ij}$ is incorporated in the our proposed method.

\subsection{Phase 1: Visited vs. Unvisited POIs}\label{sec-phase1}
In the first phase, we focus on ranking higher the POIs that a user has visited (no matter how many times) than the ones she has not visited. Formally, we aim at ranking $\mathcal{L}^+_i$ higher than $\mathcal{L}^-_i$. More specifically, our goal is to rank the POIs with emphasis on the top of the list. Moreover, with take into consideration the geographical distance between POIs. Building ranking functions that incorporate the distance between POIs also allows us to model latent associations between users living in the same neighborhood, who would not be associated with each other in a traditional CR setting. This happens because our method takes into account venues' distances as it updates the user and item latent matrices. .

Let $H_i(l_j^-)$ be the ``height" of an irrelevant venue:
\begin{equation}
    H_i(l_j^-) = \sum_{k\in\mathcal{L}_i^+} \Big(\big(\mathbf{1}_{[f_i(l_k^+) \leq f_i(l_j^-)]}\big)/\big(1 + \alpha \exp(g_{kj})\big)\Big)~,
\end{equation}

\noindent
where $\alpha$ is the weight of geographical influence and $\mathbf{1}_{[.]}$ is an indicator function. 
Note that $\alpha$ controls the model's bias towards POIs in the same neighborhood and can be used to prevent the ``Harry Potter'' problem~\cite{DBLP:conf/ecir/KoolenBBK15}.
Dividing the indicator function by $G$ allows the model to incorporate the geographical distances into the model while constructing the height for irrelevant items. For example, if an irrelevant item is ranked higher than a relevant item, but they are very close then the denominator will be higher, reducing the height of the irrelevant POI.
The objective function should aim at minimizing $H_i$ for all given POIs. A lower value of $H_i$ also means that there are fewer irrelevant POIs ranked higher than relevant ones. From an optimization perspective, indicator functions are not convex. Therefore, we use the logistic loss of the difference between the two functions as a convex upper bound surrogate. 
For reading simplicity we define
\[
  G_{\alpha}(i,j) = 1 + \alpha \exp(g_{ij})~.
\]
We define the difference between the $k^{\text{th}}$ POI and the $j^{\text{th}}$ as follows:
\begin{equation}
    \delta_i(k,j)=\mathbf{u}_i^T\big((\mathbf{v}_k-\mathbf{v}_j)/G_{\alpha}(k,j)\big)~.
\end{equation}

Therefore, the surrogate height function $H_i^{\prime}(l_j^-)$ becomes:
\begin{equation}
    H_i^{\prime}(l_j^-) = \sum_{k\in \mathcal{L}_i^+}\log\big[1+\exp\big(-\delta_i(k,j)\big)\big]~,
\end{equation}
\noindent
where $\ell(\delta) = \log(1+\exp(-\delta))$ is the logistic loss of $\delta$.
We consider $\ell_2$-norm of $H_i^{\prime}$ as the objective function, following \cite{DBLP:journals/jmlr/Rudin09}. Therefore, the objective function is:
\begin{align}
    \label{eq-obj-r}
    \begin{split}
        & R(U,V)  = \sum_{i=1}^m \frac{1}{n_i}\sum_{j\in\mathcal{L}_i^-}\big(H_i^{\prime}(l_j^-)\big)^2 = \sum_{i=1}^m \frac{1}{n_i} \times \\
         & \sum_{j\in\mathcal{L}_i^-}\Bigg( \sum_{k\in \mathcal{L}_i^+}\log\bigg(1+\exp\Big(-\mathbf{u}_i^T\big[(\mathbf{v}_k-\mathbf{v}_j)/G_{\alpha}(k,j)\big]\Big)\bigg)\Bigg)^2~.
    \end{split}    
\end{align}
\noindent For solving the above optimization problem, we use a gradient-descent-based alternating optimization algorithm. We first keep $V$ fixed and update $U$, and then keep $U$ and update $V$. Therefore, the update rules of the $t+1$ iteration are:
\begin{equation}
    \label{eq-u-phase1}
    \mathbf{u}_i^{t+1} = \mathbf{u}_i^{t} - \gamma \bigtriangledown_{\mathbf{u}_i}R(U^t,V^t), \forall i = 1,\dots,n~,
\end{equation}
\begin{equation}
    \label{eq-v-phase1}
    \mathbf{v}_j^{t+1} = \mathbf{v}_j^{t} - \gamma \bigtriangledown_{\mathbf{v}_j}R(U^{t+1},V^t), \forall j = 1,\dots,m~.
\end{equation}
\noindent For reading simplicity we define

\[
\theta(k,j)=\big(1+\exp(\delta(k,j)\big)~,
\]
\noindent and the gradients of $R(U,V)$ with respect to $\mathbf{u}_i$ and $\mathbf{v}_j$ are computed as follows:

\begin{align*}
    \begin{split}
    & \bigtriangledown_{\mathbf{u}_i}R(U,V)  \\
    & = \frac{2}{n_i} \sum_{j\in\mathcal{L}_i^-}\Big( H_i^{\prime}(l_j^-) \sum_{k \in \mathcal{L}_i^+}(\mathbf{v}_j - \mathbf{v}_k) / \big(G_{\alpha}(k,j)\theta(k,j)\big)\Big)~,
    \end{split}
\end{align*}

\begin{align*}
    \begin{split}
         &  \bigtriangledown_{\mathbf{v}_j}R(U,V) \\
         & = \sum_{i\in\mathcal{P}^-_j}\frac{2}{n_i}\sum_{h\in\mathcal{L}_i^-}\Big( H_i^{\prime}(l_h^-) \sum_{k \in \mathcal{L}_i^+} \mathbf{u}_i / \big(G_{\alpha}(k,h)\theta(k,h)\big) \Big) \\
        & - \sum_{i\in\mathcal{P}^+_j}\frac{2}{n_i}\sum_{h\in\mathcal{L}_i^-}\Big( H_i^{\prime}(l_h^-) \sum_{k \in \mathcal{L}_i^+} \mathbf{u}_i / \big(G_{\alpha}(k,h)\theta(k,h) \big) \Big)~,
    \end{split}
\end{align*}
\noindent with $\mathcal{P}^+_j$ being the set of users who have visited $l_j$ and $\mathcal{P}^-_j$ the set of users who have not visited $l_j$.

\subsection{Phase 2: Multiple vs. Single Check-ins} \label{sec-phase2}
In the second phase, we focus on ranking POIs with multiple check-ins (i.e., more-relevant POIs) higher than the ones with single check-in (i.e., relevant POIs). 
As we observed in Section \ref{sec-multi}, there is a positive correlation between the number of multiple check-ins and total number check-ins. Therefore, it is crucial to take into consideration the fact that the POIs that users have visited more often in the past are more relevant.
Formally, let $\mathcal{L}^{1+}_i = \mathcal{L}^+_i - \mathcal{L}^*_i$ be the set of POIs that the $i^{\text{th}}$ user has visited exactly once. Our goal is to rank $\mathcal{L}^*_i$ higher than $\mathcal{L}^{1+}_i$. Let $\Pi_i(l_j^*)$ be the ``reverse height" of a more-relevant venue, that is:
\begin{equation}
    \Pi_i(l_j^*) = \sum_{k\in\mathcal{L}_i^{1+}} \mathbf{1}_{[f_i(l_k^*) \leq f_i(l_j^{1+})]}~.
\end{equation}
\noindent Lower values of $\Pi_i$ mean that there are fewer relevant POIs ranked higher than the more-relevant ones. Similar to the first phase in Section~\ref{sec-phase1}, we use a logistic loss as the surrogate:
\begin{equation}
    \Pi_i^{\prime}(l_j^{1+}) = \sum_{k\in \mathcal{L}_i^*}\log\big(1+\exp(-\mathbf{u}_i^T(\mathbf{v}_k-\mathbf{v}_j))\big)~.
\end{equation} 

However, $\Pi_i^{\prime}$ is not easy to be optimized using typical ranking loss, like DCG.
Hence, we reformulate the objective functions as follows:
\[
\sum_{k\in\mathcal{L}_i^*}\log\Big(1+\Pi_i(l_k^*)\Big)~.
\]
\noindent Then, the objective function $R_{\Pi}(U,V)$ of reverse height becomes:
\begin{align}
	\label{eq-obj-rpi}
    \begin{split}
         & R_{\Pi}(U,V)  = \sum_{i=1}^m \frac{1}{n_i}\sum_{j\in\mathcal{L}_j^*}\log(1+\Pi_i^{\prime}(l_k^*)) = \sum_{i=1}^m \frac{1}{n_i} \times \\
         & \sum_{j\in\mathcal{L}_j^*}\log\bigg(1+\sum_{k\in \mathcal{L}_i^{1+}}\log\Big(1+\exp\big(-\mathbf{u}_i^T(\mathbf{v}_k-\mathbf{v}_j)\big)\Big)\bigg)~.
    \end{split}    
\end{align}

\noindent We optimize $R_{\Pi}(U,V)$ similarly to Section~\ref{sec-phase1}, that is, we first keep $V$ fixed and update $U$, then keep $U$ updating $V$. Therefore, we consider the following update rules:
\begin{equation}
    \label{eq-u-phase2}
    \mathbf{u}_i^{t+1} = \mathbf{u}_i^{t} - \gamma \bigtriangledown_{\mathbf{u}_i}R_{\Pi}(U^t,V^t), \forall i = 1,\dots,n~,
\end{equation}
\begin{equation}
    \label{eq-v-phase2}
    \mathbf{v}_j^{t+1} = \mathbf{v}_j^{t} - \gamma \bigtriangledown_{\mathbf{v}_j}R_{\Pi}(U^{t+1},V^t), \forall j = 1,\dots,m~.
\end{equation}

Similarly, the gradients are defined as follows:

\begin{align*}
    \begin{split}
       & \bigtriangledown_{\mathbf{u}_i}R_{\Pi}(U,V) \\ 
       & = \frac{1}{n_i}\sum_{j\in\mathcal{L}_i^*}\Big(\frac{1}{1+\Pi_i^{\prime}(l_j^*)} \sum_{k \in \mathcal{L}_i^{1+}} \big((\mathbf{v}_j - \mathbf{v}_k) / (1+\exp(\delta_i(k,j)))\big)\Big),
    \end{split}
\end{align*}

\begin{align*}
    \begin{split}
        & \bigtriangledown_{\mathbf{v}_j}R_{\Pi}(U,V) \\ 
        & = \sum_{i\in\mathcal{P}^{1+}_j}\frac{1}{n_i}\sum_{h\in\mathcal{L}_i^{1+}}\Big(\frac{1}{1+\Pi_i^{\prime}(l_h^*)} \sum_{k \in \mathcal{L}_i^*} \big( \mathbf{u}_i / (1+\exp(\delta_i(k,h))) \big) \Big) \\
        & - \sum_{i\in\mathcal{P}^*_j}\frac{1}{n_i}\sum_{h\in\mathcal{L}_i^{1+}}\Big(\frac{1}{1+\Pi_i^{\prime}(l_h^*)} \sum_{k \in \mathcal{L}_i^*} \big( \mathbf{u}_i / (1+\exp(\delta_i(k,h))) \big) \Big)~,
    \end{split}
\end{align*}
\noindent where we define $\mathcal{P}^{1+}_j$ the set of users who have visited $l_j$ only once and $\mathcal{P}^*_j$ the set of users who have visited $l_j$ multiple times.

\subsection{Time-Sensitive Regularizer} \label{sec-reg}
As we discussed in Section~\ref{sec-time-analysis}, the long-term temporal activity patterns of both users and POIs should be taken into account. One way to account for the activity patterns of users and the popularity of POIs is to consider how variant they are over time. For example, if a POI is a coffee shop and receives approximately the same number of people every month, it is more likely that it receives the same number of users in the next month. Whereas, POIs like ski shop are only popular during the ski season.
In particular, we observed in Figure \ref{fig-vars-most-v} that the popularity of certain POI categories are highly time-dependent.
Here, we propose to incorporate a novel time-sensitive regularizer into the objective function of both of our objective phases in (\ref{eq-obj-r}) and (\ref{eq-obj-rpi}). Adding a regularizer that is calculated for each user and POI based on their past activities enables us to model the time-sensitivity of users and POIs. The main goal here is to penalize those users and POIs which are less stable. A more stable user or POI is one that exhibits less activity variance over time. This regularizer is defined based on our extensive analysis and observation in Section~\ref{sec-time-analysis} where we observed that POIs that are less popular are more time-sensitive.  We also had a similar observation for users, where we observed that less active users exhibit less stability in their check-in behavior. 

Let $\sigma^{2,U} \in \mathbb{R}^{1\times n}$ be the variance vector for users, where $\sigma^{2,U}_i$ is the activity variance of the $i^{\text{th}}$ user $\rho_i$. 
For each user $\rho_i$, we count the number of check-ins per month and normalize the values. Then, calculating the variance of the monthly check-ins of $\rho_i$ produces  $\sigma^{2,U}_i$.
Similarly, let $\sigma^{2,V} \in \mathbb{R}^{1\times m}$ be the variance vector for POIs, with $\sigma^{2,V}_j$ being the popularity variance of the $j^{\text{th}}$ POI $l_j$. Note that we calculate the variance of POIs based on their corresponding categories since we observed more meaningful popularity patterns with respect to the categories. The time-sensitive regularizer parameter for users and POIs are calculated as follows:
\begin{equation}
    \Lambda^U = \lambda\log\big(1+\exp(-\sigma^{2,U})\big)~,
\end{equation}
\begin{equation}
    \Lambda^V = \lambda\log\big(1+\exp(-\sigma^{2,V})\big)~.
\end{equation}

It is worth noting that we consider the logistic function of variances to prevent underflow and take the hyper parameter $\lambda$ as a controlling parameter to prevent the model from overfitting. Ultimately, we add the time-sensitive regularizer to the objective functions of our two phases in (\ref{eq-obj-r}) and (\ref{eq-obj-rpi}). Thus, when updating $\mathbf{u}_i$, we add the regularizer term $\Lambda^U_i\mathbf{u}_i$, and for updating $\mathbf{v_j}$ we add the regularizer term $\Lambda^V_j\mathbf{v}_j$.

\begin{algorithm}[pt]
\DontPrintSemicolon
\KwIn{ $\mathcal{P}$, $\mathcal{L}$, $maxIter$, $\{\lambda, \gamma, \alpha, \epsilon, d\}$\\
}
\KwOut{$U_{\text{out}}$, $V_{\text{out}}$ \\
}
$t \gets 0$ \\
Initialize $U^{t+1}$, $V^{t+1}$ \\
$\theta^{t+1} \gets R(U^{t+1},V^{t+1}) + R_\Pi(U^{t+1}, V^{t+1})$, $\theta^t = \frac{\theta^{t+1}}{2}$ \\

\While{($abs(\theta^{t+1} - \theta^{t}) > \epsilon$) $\land$ ($t < maxIter$)} {
    \quad $t \gets t+1$ \\
    
    \quad \tcp{Phase 1} 
	\quad Update $\mathbf{u}_i^{t+1}, \forall i = 1,\dots,n$ Eq. (\ref{eq-u-phase1}) \\
	\quad Update $\mathbf{v}_j^{t+1}, \forall j = 1,\dots,m$ Eq. (\ref{eq-v-phase1}) \\	
    \quad \tcp{Phase 2} 
    \quad Update $\mathbf{u}_i^{t+1}, \forall i = 1,\dots,n$ Eq. (\ref{eq-u-phase2}) \\
	\quad Update $\mathbf{v}_j^{t+1}, \forall j = 1,\dots,m$ Eq. (\ref{eq-v-phase2}) \\	
	
	\quad $\theta^{t+1} \gets R(U^{t+1},V^{t+1}) + R_\Pi(U^{t+1}, V^{t+1})$ \\
\nonl\bf{end}
}

$U_{\text{out}} \gets U^{t+1}, V_{\text{out}} \gets V^{t+1}$ \\

\caption{{The Joint Two-Phase Collaborative Ranking Algorithm (\modelname).}}
\label{algo-tscr}
\end{algorithm}

\subsection{Joint Two-Phase Collaborative Ranking Algorithm} \label{sec-opt}
Algorithm~\ref{algo-tscr} presents the proposed joint two-phase collaborative ranking method.
Line 2 initializes the factor matrices randomly. $\theta$ is initialized at line 3, summing up the values of the two phases of our objective function, namely, $R(U^{t+1},V^{t+1})$ and $R_\Pi(U^{t+1}, V^{t+1})$. The joint optimization of the two phases is done between lines 4 and 12. As we see, in every iteration, $\mathbf{u}_i$ and $\mathbf{v}_j$ are first updated according to \eqref{eq-u-phase1} and \eqref{eq-v-phase1} (lines 7 and 8) to push visited POIs higher in the ranking. Each iteration is then followed by optimizing $\mathbf{u}_i$ and $\mathbf{v}_j$ according to \eqref{eq-u-phase2} and \eqref{eq-v-phase2} (lines 10 and 11), respectively. Therefore, $U$ and $V$ factor matrices are optimized jointly to push visited POIs higher than unvisited POIs and multiple visited POIs higher than single visited POIs simultaneously. 
After convergence, the final values of the latent factor matrices are stored at line 13. Note that the proposed time-sensitive regularizer is applied at lines 7, 8, 10, and 11. Also, the geographical influence is applied at lines 7 and 8.
One can argue that employing a two-phase learning strategy might be computationally expensive. However, since in the second phase we only focus on the POIs that each user has checked in, the optimization algorithm does not add a substantial overhead to the whole system. In fact, for each user the complexity of one iteration is $\mathcal{O}(n_i^{1+}n_i^*)$, which is very small compared to the first phase, which is $\mathcal{O}(n_i^-n_i^+)$.

\section{Experimental Evaluation}\label{sec-experiments}

In this section, we evaluate the performance of our model compared with state-of-the-art methods and study the impact of different parameters on the performance of our model.

\subsection{Experimental Setup and Metrics}
We evaluate our method on two real-world datasets, namely, Foursquare's and Gowalla's. Both datasets were provided by the authors of~\cite{DBLP:conf/sigir/YuanCMSM13}. The statistical details of the datasets are listed in \autoref{tab-stats}. We take the first 70\% of the data for each user as the training set, 10\% as the validation set, and the remaining 20\% as the test set, following the evaluation protocol of \cite{DBLP:conf/www/ZhaoZKL17}.

We compare the performance of our model in terms of \textit{Precision at k} (Prec@k) and \textit{Normalized Discounted Cumulative Gain at k} (nDCG@k). 
Let $L_{ch}(\rho)$ be the set POIs that a user has visited in the test set and $L_{rec}^{k}(\rho)$ be the set of recommended POIs of size $k$. Prec@k$(\rho)$ for a user $\rho$ is defined as
$\text{Prec@k}(\rho) = (|L_{ch}(\rho) \cap L_{rec}^k(\rho)|)/(k)$ 
and Prec@k for the whole dataset is the average Prec@k$(\rho)$ for all the users in the test set. 

To calculate nDCG@k, we need to define relevance values in the test set. Following the same strategy of Section~\ref{sec-method}, we define a three-level relevance for each POI based on the frequency of check-in for a particular user:
\[ rel(l,\rho) =
  \begin{cases}
    2  & \quad \text{if } \rho \text{ visited } l \text{ multiple times}\\
    1  & \quad \text{if } \rho \text{ visited } l \text{ only once}\\
    0  & \quad \text{if } \rho \text{ did not visit } l
  \end{cases}~.
\]
Therefore, nDCG@k$(\rho)$ for a given user $\rho$ is defined as follows:
\[
\text{DCG@k}(\rho) = \sum_{r=1}^k \frac{2^{rel(l_r,\rho)}-1}{\log_2(r+1)},
\]
\[
\text{nDCG@k}(\rho) = \frac{\text{DCG@k}(\rho)}{\text{IDCG@k}(\rho)}~,
\]
where $l_r$ is the POI at the $r^{\text{th}}$ rank and IDCG@k$(\rho)$ is the ideal DCG@k value for user $\rho$, that is, the highest possible value for DCG@k. The reported values of nDCG@k are the average of the nDCG@k$(\rho)$ values for all the users in the test set. We report the values of nDCG@k and Prec@k for three values of k, namely 5, 10, 20.

\subsection{Compared Methods}
We compare our \textbf{J}oint \textbf{T}wo-Phase \textbf{C}ollaborative \textbf{R}anking (\modelname) model with approaches that consider geographical influence for POI recommendation and approaches based on collaborative ranking with emphasis on ranking relevant items higher. Also, we include two variations of the proposed \modelname to demonstrate the effectiveness of different elements of our algorithm. Note that for each model, we find the optimum set of parameters using the validation set and report the mean and standard deviation of results of 5 different runs with the same parameters.
We compare our \modelname model with the following methods:

\begin{itemize}
    \item \textbf{\modelname-Phase1} reports the performance of the first phase of \modelname. We include this model as a baseline to demonstrate the effectiveness of the first phase of \modelname and the significance of the second phase of the algorithm.
    \item \textbf{\modelname-NoVar} reports the result of our proposed \modelname without using the time-sensitive regularizer. Instead, we use $\lambda / 2 (\|U\|^2+\|V\|^2)$  as the regularizer. Our goal is to demonstrate the effectiveness of the time-based regularizer.
    \item \textbf{\modelname-NoGeo} reports the result of our proposed \modelname without applying the geographical influence (i.e., $\alpha=0$).
    \item \textbf{WRMF}~\cite{DBLP:conf/icdm/HuKV08} proposes an MF method for item prediction from implicit feedback. It is an adaptation of SVD, minimizing the square-loss.
	\item \textbf{IRenMF}~\cite{DBLP:conf/cikm/LiuWSM14} is based on weighted MF~\cite{DBLP:conf/icdm/PanZCLLSY08} exploiting two levels of geographical neighborhood characteristics: nearest neighboring locations share more similar user preferences, while locations in the same geographical region may share similar user preferences.
    \item \textbf{GeoMF}~\cite{DBLP:conf/kdd/LianZXSCR14} augments users’ and venues’ latent factors in the factorization model with activity area vectors of
users and influence area vectors of venues, respectively.
    \item \textbf{Rank-GeoFM}~\cite{DBLP:conf/sigir/LiCLPK15} is a ranking-based MF model that includes the geographical influence of neighboring venues while learning users’ preference rankings for venues.
    \item \textbf{Rank-GeoFM-NoGeo} reports the result of Rank-GeoFM without considering the geographical influence.
    \item \textbf{RH-Push / Inf-Push / P-Push}~\cite{DBLP:conf/www/Christakopoulou15} are three push CR models based on reverse height, infinite, and $p$-norm. For each user, we considered the venues she visited as positive training samples and selected $k$ venues randomly as negative training samples. 
\end{itemize}

We aim to compare the performance of \modelname against state-of-the-art methods in POI recommendation which consider recommendation as a ranking problem and the ones that do not. Also, it is crucial to compare our method with approaches that incorporate geographical influence into the model. The other set of methods is based on CR. Our aim is to demonstrate the effectiveness of our two-phase regularized CR in comparison with other CR baselines.

\begin{table*}
\caption{Performance evaluation on Foursquare's in terms of nDCG@k. Statistically significant differences with \modelname are denoted by $\smalltriangledown$ for $p<$0.05 in paired t-test. $\Delta$ values express the relative difference, compared to \modelname. For each model we report the mean and standard deviation of 5 different runs.}
\label{tab-results-fsq-ndcg}
\centering
\setlength{\extrarowheight}{.3em}
\begin{tabular}{l@{\quad}l@{\quad}l@{\quad}l@{\quad}l@{\quad}l@{\quad}l@{\quad}l@{\quad}l}
\toprule
& \multicolumn{1}{c}{\textbf{nDCG@5}} & \multicolumn{1}{c}{$\Delta$} && \multicolumn{1}{c}{\textbf{nDCG@10}} & \multicolumn{1}{c}{$\Delta$} && \multicolumn{1}{c}{\textbf{nDCG@20}} & \multicolumn{1}{c}{$\Delta$} \\ 
\cmidrule(lr){2-3}  \cmidrule(rl){4-6} \cmidrule(rl){7-9}
\textbf{\modelname} & $0.0639 \pm 0.0015$ & \multicolumn{1}{c}{-} &&  $0.0529 \pm 0.0033$ & \multicolumn{1}{c}{-} && $0.0394 \pm 0.0019$ & \multicolumn{1}{c}{-} \\
        \textbf{\modelname-Phase1} & $0.0534 \pm 0.0022^\smalltriangledown$ & $-16.43\%$ &&  $0.0436 \pm 0.0019^\smalltriangledown$ & $-17.58\%$ && $0.0339 \pm 0.0027^\smalltriangledown$ & $-13.96\%$ \\
        \textbf{\modelname-NoVar} & $0.0605 \pm 0.0025^\smalltriangledown$ & $-5.32\%$ &&  $0.0497 \pm 0.0043^\smalltriangledown$ & $-6.05\%$ && $0.0375 \pm 0.0036^\smalltriangledown$ & $-4.82\%$ \\
        \textbf{\modelname-NoGeo} & $0.0613 \pm 0.0017^\smalltriangledown$ & $-4.07\%$ && $0.0494 \pm 0.0022^\smalltriangledown$ & $-6.62\%$ && $0.0381 \pm 0.0019$ & $-3.3\%$ \\
        \textbf{WRMF} & $0.0248 \pm 0.0016^\smalltriangledown$ & $-61.19\%$ && $0.0210 \pm 0.0014^\smalltriangledown$ & $-60.3\%$ && $0.0178 \pm 0.0025^\smalltriangledown$ & $-54.82\%$ \\
        \textbf{GeoMF} & $0.0422 \pm 0.0018^\smalltriangledown$ & $-33.96\%$ &&  $0.0336 \pm 0.0016\smalltriangledown$ & $-36.48\%$ && $0.0250 \pm 0.0021^\smalltriangledown$ & $-36.55\%$ \\
        \textbf{IRenMF} & $0.0430 \pm 0.0011^\smalltriangledown$ & $-32.71\%$ &&  $0.0348 \pm 0.0009^\smalltriangledown$ & $-34.22\%$ && $0.0286 \pm 0.0010^\smalltriangledown$ & $-27.41\%$ \\
        \textbf{Rank-GeoFM} & $0.0438 \pm 0.0005^\smalltriangledown$ & $-31.46\%$ &&  $0.0359 \pm 0.0001^\smalltriangledown$ & $-32.14\%$ && $0.0277 \pm 0.0001^\smalltriangledown$ & $-29.70\%$ \\
        \textbf{Rank-GeoFM-NoGeo} & $0.0418 \pm 0.0018^\smalltriangledown$ & $-34.59\%$ && $0.0314 \pm 0.0012^\smalltriangledown$ & $-40.64\%$ && $0.0234 \pm 0.0005^\smalltriangledown$ & $-40.61\%$ \\
        \textbf{RH-Push} & $0.0251 \pm 0.0044^\smalltriangledown$ & $-60.72\%$ &&  $0.0187 \pm 0.0040^\smalltriangledown$ & $-64.65\%$ && $0.0137 \pm 0.0039^\smalltriangledown$ & $-65.23\%$ \\
        \textbf{Inf-Push} & $0.0433 \pm 0.0053^\smalltriangledown$ & $-32.24\%$ &&  $0.0361 \pm 0.0051^\smalltriangledown$ & $-31.76\%$ && $0.0302 \pm 0.0054^\smalltriangledown$ & $-23.35\%$ \\
        \textbf{P-Push} & $0.0423 \pm 0.0068^\smalltriangledown$ & $-33.8\%$ &&  $0.0309 \pm 0.0065^\smalltriangledown$ & $-41.59\%$ && $0.0218 \pm 0.0071^\smalltriangledown$ & $-44.67\%$ \\
\bottomrule
\end{tabular}

\bigskip

\caption{Performance evaluation on Foursquare's in terms of Prec@k. Statistically significant differences with \modelname are denoted by $\smalltriangledown$ for $p<$0.05 in paired t-test. $\Delta$ values express the relative difference, compared to \modelname. For each model we report the mean and standard deviation of 5 different runs.}
\label{tab-results-fsq-prec}
\centering
\setlength{\extrarowheight}{.3em}
\begin{tabular}{l@{\quad}l@{\quad}l@{\quad}l@{\quad}l@{\quad}l@{\quad}l@{\quad}l@{\quad}l}
\toprule
& \multicolumn{1}{c}{\textbf{Prec@5}} & \multicolumn{1}{c}{$\Delta$} && \multicolumn{1}{c}{\textbf{Prec@10}} & \multicolumn{1}{c}{$\Delta$} && \multicolumn{1}{c}{\textbf{Prec@20}} & \multicolumn{1}{c}{$\Delta$} \\ 
\cmidrule(lr){2-3}  \cmidrule(rl){4-6} \cmidrule(rl){7-9}
				
        \textbf{\modelname} & $0.0591 \pm 0.0025$ & \multicolumn{1}{c}{-} &&  $0.0456 \pm 0.0041$ & \multicolumn{1}{c}{-} && $0.0303 \pm 0.0046$ & \multicolumn{1}{c}{-} \\        
        
        \textbf{\modelname-Phase1} & $0.0462 \pm 0.0013^\smalltriangledown$ & $-21.83\%$ &&  $0.0357 \pm 0.0022^\smalltriangledown$ & $-21.71\%$ && $0.0260 \pm 0.0021^\smalltriangledown$ & $-14.19\%$ \\
        \textbf{\modelname-NoVar} & $0.0536 \pm 0.0018^\smalltriangledown$ & $-9.31\%$ &&  $0.0414 \pm 0.0015^\smalltriangledown$ & $-9.21\%$ && $0.0282 \pm 0.0025^\smalltriangledown$ & $-6.93\%$ \\
        \textbf{\modelname-NoGeo} & $0.0551 \pm 0.0032^\smalltriangledown$ &	$-6.77\%$ && $0.0411 \pm 0.0027^\smalltriangledown$ & $-9.87\%$ && $0.0292 \pm 0.0018$ & $-3.63\%$\\
        \textbf{WRMF} & $0.0224 \pm 0.0020^\smalltriangledown$ & $-62.1\%$ && $0.0181 \pm 0.0012^\smalltriangledown$ & $-60.31\%$ && $0.0151 \pm 0.0017^\smalltriangledown$ & $-50.17\%$\\
        \textbf{GeoMF} & $0.0385 \pm 0.0015^\smalltriangledown$ & $-34.86\%$ &&  $0.0281 \pm 0.0016^\smalltriangledown$ & $-38.38\%$ && $0.0186 \pm 0.0012^\smalltriangledown$ & $-38.61\%$ \\
        \textbf{IRenMF} & $0.0385 \pm 0.0009^\smalltriangledown$ & $-34.86\%$ &&  $0.0280 \pm 0.0011^\smalltriangledown$ & $-38.60\%$ && $0.0219 \pm 0.0015^\smalltriangledown$ & $-27.72\%$ \\
        \textbf{Rank-GeoFM} & $0.0397 \pm 0.0004^\smalltriangledown$ & $-32.83\%$ &&  $0.0304 \pm 0.0002^\smalltriangledown$ & $-33.33\%$ && $0.0215 \pm 0.0001^\smalltriangledown$ & $-29.04\%$ \\
        \textbf{Rank-GeoFM-NoGeo} & $0.0339 \pm 0.0007^\smalltriangledown$ & $-42.64\%$ && $0.0229 \pm 0.0003^\smalltriangledown$ & $-49.78\%$ && $0.0157 \pm 0.0003^\smalltriangledown$ & $-48.18\%$ \\
        \textbf{RH-Push} & $0.0214 \pm 0.0042^\smalltriangledown$ & $-63.79\%$ &&  $0.0140 \pm 0.0051^\smalltriangledown$ & $-69.3\%$ && $0.0094 \pm 0.0046^\smalltriangledown$ & $-68.98\%$ \\
        \textbf{Inf-Push} & $0.0397 \pm 0.0051^\smalltriangledown$ & $-32.83\%$ &&  $0.0313 \pm 0.0061^\smalltriangledown$ & $-31.36\%$ && $0.0217 \pm 0.0066^\smalltriangledown$ & $-28.38\%$ \\
        \textbf{P-Push} & $0.0326 \pm 0.0072^\smalltriangledown$ & $-44.84\%$ &&  $0.0211 \pm 0.0101^\smalltriangledown$ & $-53.73\%$ && $0.0132 \pm 0.0094^\smalltriangledown$ & $-56.44\%$ \\

\bottomrule
\end{tabular}
\end{table*}

\begin{table*}
\caption{Performance evaluation on Gowalla's in terms of nDCG@k. Statistically significant differences with \modelname are denoted by $\smalltriangledown$ for $p<$0.05 in paired t-test. $\Delta$ values express the relative difference, compared to \modelname. For each model we report the mean and standard deviation of 5 different runs.}
\label{tab-results-gwl-ndcg}
\centering
\setlength{\extrarowheight}{.3em}
\begin{tabular}{l@{\quad}l@{\quad}l@{\quad}l@{\quad}l@{\quad}l@{\quad}l@{\quad}l@{\quad}l}
\toprule
& \multicolumn{1}{c}{\textbf{nDCG@5}} & \multicolumn{1}{c}{$\Delta$} && \multicolumn{1}{c}{\textbf{nDCG@10}} & \multicolumn{1}{c}{$\Delta$} && \multicolumn{1}{c}{\textbf{nDCG@20}} & \multicolumn{1}{c}{$\Delta$} \\ 
\cmidrule(lr){2-3}  \cmidrule(rl){4-6} \cmidrule(rl){7-9}
        % \textbf{\modelname} & $0.1143 \pm 0.0018$ & \multicolumn{1}{c}{-} &&  $0.0848 \pm 0.0004$ & \multicolumn{1}{c}{-} && $0.0625 \pm 0.0004$ & \multicolumn{1}{c}{-} \\
        \textbf{\modelname} & $0.1158 \pm 0.0022$ & \multicolumn{1}{c}{-}&&  $0.0854 \pm 0.0007$ & \multicolumn{1}{c}{-} && $0.0633 \pm 0.0008$ & \multicolumn{1}{c}{-} \\     
        \textbf{\modelname-Phase1} & $0.1090 \pm 0.0021^\smalltriangledown$ & $-4.32\%$ &&  $0.0823 \pm 0.0015$ & $-3.04\%$ && $0.0607 \pm 0.0012$ & $-2.53\%$ \\
        \textbf{\modelname-NoVar} & $0.1092 \pm 0.0029^\smalltriangledown$ & $-5.79\%$ &&  $0.0802 \pm 0.0022^\smalltriangledown$ & $-5.85\%$ && $0.0593 \pm 0.0009^\smalltriangledown$ & $-7.42\%$ \\
        \textbf{\modelname-NoGeo} & $0.1099 \pm 0.0015^\smalltriangledown$ & $-5.09\%$ && $0.0823 \pm 0.0010$ & $-3.63\%$ && $0.0608 \pm 0.0006$ & $-3.95^\smalltriangledown\%$ \\
        \textbf{WRMF} & $0.0620 \pm 0.0023^\smalltriangledown$ & $-46.46\%$ && $0.0523 \pm 0.0011^\smalltriangledown$ & $-38.76\%$ && $0.0425 \pm 0.0012^\smalltriangledown$ & $-32.86\%$ \\
        \textbf{GeoMF} & $0.0604 \pm 0.0031^\smalltriangledown$ & $-47.84\%$ &&  $0.0495 \pm 0.0037^\smalltriangledown$ & $-42.04\%$ && $0.0374 \pm 0.0031^\smalltriangledown$ & $-40.92\%$ \\
        \textbf{IRenMF} & $0.0606 \pm 0.0043^\smalltriangledown$ & $-47.67\%$ &&  $0.0499 \pm 0.0039^\smalltriangledown$ & $-41.57\%$ && $0.0389 \pm 0.0025^\smalltriangledown$ & $-38.55\%$ \\
        \textbf{Rank-GeoFM} & $0.0593 \pm 0.0008^\smalltriangledown$ & $-48.79\%$ &&  $0.0525 \pm 0.0006^\smalltriangledown$ & $-38.52\%$ && $0.0451 \pm 0.0004^\smalltriangledown$ & $-28.75\%$ \\
        \textbf{Rank-GeoFM-NoGeo} & $0.0675 \pm 0.0011^\smalltriangledown$ & $-41.71\%$ && $0.0514 \pm 0.0009^\smalltriangledown$ & $-39.81\%$ && $0.0387 \pm 0.0008^\smalltriangledown$ & $-38.86\%$ \\
        \textbf{RH-Push} & $0.0985 \pm 0.0031^\smalltriangledown$ & $-14.94\%$ &&  $0.0765 \pm 0.0026^\smalltriangledown$ & $-10.42\%$ && $0.0569 \pm 0.0037^\smalltriangledown$ & $-10.11\%$ \\
        \textbf{Inf-Push} & $0.1090 \pm 0.0035^\smalltriangledown$ & $-5.87\%$ &&  $0.0803 \pm 0.0028^\smalltriangledown$ & $-5.97\%$ && $0.0585 \pm 0.0018^\smalltriangledown$ & $-7.58\%$ \\
        \textbf{P-Push} & $0.1026 \pm 0.0029^\smalltriangledown$ & $-11.40\%$ &&  $0.0805 \pm 0.0013^\smalltriangledown$ & $-5.74\%$ && $0.0596 \pm 0.0015^\smalltriangledown$ & $-5.85\%$ \\
\bottomrule
\end{tabular}

\bigskip

\caption{Performance evaluation on Gowalla's in terms of Prec@k. Statistically significant differences with \modelname are denoted by $\smalltriangledown$ for $p<$0.05 in paired t-test. $\Delta$ values express the relative difference, compared to \modelname. For each model we report the mean and standard deviation of 5 different runs.}
\label{tab-results--gwl-prec}
\centering
\setlength{\extrarowheight}{.3em}
\begin{tabular}{l@{\quad}l@{\quad}l@{\quad}l@{\quad}l@{\quad}l@{\quad}l@{\quad}l@{\quad}l}
\toprule
& \multicolumn{1}{c}{\textbf{Prec@5}} & \multicolumn{1}{c}{$\Delta$} && \multicolumn{1}{c}{\textbf{Prec@10}} & \multicolumn{1}{c}{$\Delta$} && \multicolumn{1}{c}{\textbf{Prec@20}} & \multicolumn{1}{c}{$\Delta$} \\ 
\cmidrule(lr){2-3}  \cmidrule(rl){4-6} \cmidrule(rl){7-9}
        \textbf{\modelname} & $0.0949 \pm 0.0021$ & \multicolumn{1}{c}{-}&&  $0.0621 \pm 0.0007$ & \multicolumn{1}{c}{-} && $0.0425 \pm 0.0007$ & \multicolumn{1}{c}{-} \\     
       
        \textbf{\modelname-Phase1} & $0.0889 \pm 0.0018^\smalltriangledown$ & $-6.32\%$ &&  $0.0596 \pm 0.0012^\smalltriangledown$ & $-4.03\%$ && $0.0414 \pm 0.0011$ & $-2.59\%$ \\
        \textbf{\modelname-NoVar} & $0.0865 \pm 0.0032^\smalltriangledown$ & $-8.85\%$ &&  $0.0570 \pm 0.0015^\smalltriangledown$ & $-8.21\%$ && $0.0378 \pm 0.0012^\smalltriangledown$ & $-11.06\%$ \\
        \textbf{\modelname-NoGeo} & $0.0866 \pm 0.0022^\smalltriangledown$ & $-8.75\%$ && $0.0590 \pm 0.0020^\smalltriangledown$ & $-4.99\%$ && $0.0401 \pm 0.0012^\smalltriangledown$ & $-5.65\%$ \\
        \textbf{WRMF} & $0.0556 \pm 0.0018^\smalltriangledown$ & $-41.41\%$ && $0.0448 \pm 0.0012^\smalltriangledown$ & $-27.86\%$ && $0.0346 \pm 0.0010^\smalltriangledown$ & $-18.59\%$ \\
        \textbf{GeoMF} & $0.0540 \pm 0.0028^\smalltriangledown$ & $-43.10\%$ &&  $0.0415 \pm 0.0042^\smalltriangledown$ & $-33.17\%$ && $0.0284 \pm 0.0032^\smalltriangledown$ & $-33.18\%$ \\
        \textbf{IRenMF} & $0.0545 \pm 0.0051^\smalltriangledown$ & $-42.57\%$ &&  $0.0423 \pm 0.0043^\smalltriangledown$ & $-31.88\%$ && $0.0305 \pm 0.0023^\smalltriangledown$ & $-28.24\%$ \\
        \textbf{Rank-GeoFM} & $0.0564 \pm 0.0009^\smalltriangledown$ & $-40.57\%$ &&  $0.0472 \pm 0.0008^\smalltriangledown$ & $-23.99\%$ && $0.0384 \pm 0.0004^\smalltriangledown$ & $-9.65\%$ \\
        \textbf{Rank-GeoFM-NoGeo} & $0.0549 \pm 0.0008^\smalltriangledown$ & $-42.15\%$ && $0.0383 \pm 0.0008^\smalltriangledown$ & $-38.33\%$ && $0.0269 \pm 0.0007^\smalltriangledown$ & $-36.71\%$ \\
        
        \textbf{RH-Push} & $0.0820 \pm 0.0039^\smalltriangledown$ & $-13.59\%$ &&  $0.0588 \pm 0.0030^\smalltriangledown$ & $-5.31\%$ && $0.0400 \pm 0.0042^\smalltriangledown$ & $-5.88\%$ \\
        \textbf{Inf-Push} & $0.0864 \pm 0.0034^\smalltriangledown$ & $-8.96\%$ &&  $0.0569 \pm 0.0026^\smalltriangledown$ & $-8.37\%$ && $0.0377 \pm 0.0016^\smalltriangledown$ & $-11.29\%$ \\
        \textbf{P-Push} & $0.0844 \pm 0.0029^\smalltriangledown$ & $-11.06\%$ &&  $0.0573 \pm 0.0012^\smalltriangledown$ & $-7.73\%$ && $0.0396 \pm 0.0018^\smalltriangledown$ & $-6.82\%$ \\
        
\bottomrule
\end{tabular}
\end{table*}

\subsection{Performance Comparison} 
Tables \ref{tab-results-fsq-ndcg}, \ref{tab-results-fsq-prec}, \ref{tab-results-gwl-ndcg}, and \ref{tab-results--gwl-prec} report the performance of our method compared with 11 baselines in terms of nDCG@k and Prec@k for Foursquare's and Gowalla's. Based on the results we observe that our proposed \modelname significantly outperforms all the baseline methods on both datasets with respect to both nDCG@k and Prec@k. It is worth noting that the improvement is achieved for all values of k.

Moreover, Rank-GeoFM performs best among the geographical-based methods, as Rank-GeoFM propagates geographical influences using the constructed graph, which confirms that geographical neighborhood is a major factor for recommendation. Rank-GeoFM considers the implicit feedback while training the model similar to us, however, as we observe our two-phase collaborative ranking approach beats Rank-GeoFM indicating the effectiveness of our approach. Moreover, \modelname outperforms Rank-GeoFM by a large margin on both datasets. 
It is worth noting that \modelname beats all geographical-based methods in terms of both nDCG@k and Prec@k for all different values of k in both datasets. GeoMF and IRenMF do not consider POI recommendation as a ranking problem. Hence, they attempt to optimize the overall error rate which proves to be less effective for POI recommendation mainly because the users are only interested in top k recommended POIs. Consequently, \modelname beats GeoMF and IRenMF with a large margin.
   
In addition, we observe that in most cases CR-based baseline methods, namely P-Push, Inf-Push, and RH-Push perform better than other baseline methods. This suggests that a CR-based approach leads to better performance for POI recommendation in general.
 However, we observe that \modelname outperforms all CR-based baselines. This indicates that all CR-based methods suffer from the sparsity problem while our two-phase CR strategy alleviates this problem by considering both visited and unvisited POIs in the same neighborhood.
Also, none of the CR-based methods consider time in their ranking loss function. While it is important to consider POI recommendation as a ranking problem, it is also important to consider time to generate accurate recommendations (Section~\ref{sec-analysis}). Our proposed model beats the CR-based methods by a significant margin showing out time-sensitive regularizer (Section~\ref{sec-reg}) based on the temporal behavior of users and the temporal popularity of POIs leads to a more accurate performance.
Although the CR-based baseline approaches focus on the top of the ranked list, they fail to rank more-relevant POIs higher in the ranking. In fact, these methods consider a binary relevance between users and POIs. Higher nDCG@k values indicate that our model ranks POIs with multiple check-ins more accurately, compared with the CR-based baseline approaches. This suggests that our two-phase model ranks the POIs with higher relevance more effectively than the CR approaches by considering multi-level implicit user feedback. It is worth noting that the variants of our model also outperform most of the baselines. In particular, \modelname-NoVar outperforms all the baselines and \modelname-Phase1 performs better than most of the baselines, including P-Push. This is important, since it indicates that incorporating the temporal information together with geographical similarities improves the performance of this model when the first phase is only considered. Also, we see that \modelname-NoGeo beats all the baselines. Specifically, it performs better that Rank-GeoFM-NoGeo and other baselines that do not consider geographical information. This indicates the effectiveness of the proposed model even when the geographical influence is not considered.

\subsection{Effect of the 2$^\text{nd}$ Phase}
To study the effect of the second phase of \modelname, we compare the performance of \modelname when only the first phase is used (\modelname-Phase1) with the performance of \modelname when both phases are considered. As we can see in Tables \ref{tab-results-fsq-ndcg} - \ref{tab-results--gwl-prec}, \modelname exhibits a significant improvement over \modelname-Phase1 in terms of all evaluation metrics for both datasets. This indicates that while \modelname-Phase1 is able to beat all other baselines, the second phase of the algorithm enables \modelname to model multiple check-ins more accurately. This validates the remark based on our analysis in Section~\ref{sec-intuitions} which states that a user who has visited a POI multiple times in the past is likely to visit the same venue in the future. Moreover, while this remark based on our analysis applies to a user, it is also valid with respect to similar users. Therefore, in the second phase, similar users and their corresponding collaborative associations are mainly determined based on how similar they are in terms of multiple check-ins. This helps the model rank ``more relevant" items higher and hence improves the accuracy of the model.
Moreover, we observe a higher relative difference on Foursquare's. According to \autoref{tab-stats}, Foursquare's consists of more multiple check-ins than Gowalla's ($45.51\%$ as opposed to $32.69\%$) which explains why the second phase of \modelname achieves higher improvement on Foursquare's. 

\subsection{Effect of the Time-Sensitive Regularizer}
Next, we discuss the effect of the time-sensitive regularizer. To this end, we compare the performance of \modelname without using the time-sensitive regularizer (\modelname-NoVar). 
As seen in Tables \ref{tab-results-fsq-ndcg} \& \ref{tab-results-fsq-prec}, a statistically significant improvement of \modelname over \modelname-NoVar is observed in terms of all evaluation metrics for Foursquare's, suggesting that using the time-sensitive regularizer enables \modelname to place more relevant venues higher in the ranking. As for Gowalla's, we also see significant improvements in Tables~\ref{tab-results-gwl-ndcg} \& \ref{tab-results--gwl-prec} indicating that our proposed time-sensitive regularizer improves the performance of \modelname by penalizing those users and POIs that exhibit less stability in their check-in and popularity, respectively. This validates the remark based on our analysis that we had in Section~\ref{sec-intuitions} where we showed that there is a negative correlation between a POI's popularity and its popularity variance. A similar observation was made for users. Based on this remark, we defined the time-sensitive regularizer to penalize those users and POIs that have been variant in the past. In other words, variant users are less probable to visit variant POIs and the introduced regularizer enables the model to take this into account while training.

\subsection{Effect of the Geographical Influence}
Here, we discuss the effect of the geographical influence. Therefore, we compare the performance of \modelname without applying the geographical influence (\modelname-NoGeo). 
As seen in Tables \ref{tab-results-fsq-ndcg} \& \ref{tab-results-fsq-prec}, we observe a statistically significant improvement of \modelname over \modelname-NoGeo in terms of all evaluation metrics for Foursquare's except for nDCG@20 and Prec@20. However, the significant improvement in terms of other evaluation metrics for Foursquare's suggests that applying the geographical influence enables \modelname to model users' geographical behavior and activities more effectively. As for Gowalla's, we see significant improvements in Tables~\ref{tab-results-gwl-ndcg} \& \ref{tab-results--gwl-prec} for all evaluation metrics expect nDCG@10 and nDCG@20. This indicates that the geographical influence improves the performance of \modelname by considering how users like POIs that are in the same neighborhood.

\subsection{Effect of the Model Parameters} 
We demonstrate the effect of the model's parameters. The results reported in the previous sections are achieved after the best parameter set was found on the validation set. We fixed the learning rate ($\gamma=\num{1e-4}$) for both datasets to ensure the generalization of our model.

\begin{figure}[t]
	\begin{subfigure}[b]{0.49\columnwidth}
        \includegraphics[height=0.15\textheight]{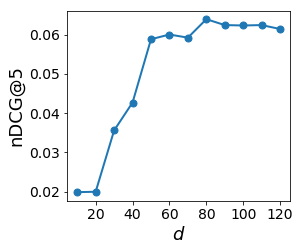}
        \caption{Foursquare's}
        \label{fig-dim-fsq}
    \end{subfigure}
	\begin{subfigure}[b]{0.49\columnwidth}
        \includegraphics[height=0.15\textheight]{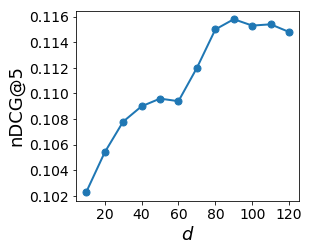}
        \caption{Gowalla's}
        \label{fig-dim-gwl}
    \end{subfigure}
\caption{Effect of the number of latent factors.} 
\label{fig-dim}
\end{figure}

In \autoref{fig-dim} we study the effect of latent factors $d$ on the performance of our model and report nDCG@5 while keeping other parameters fixed.
As we can see in \autoref{fig-dim-fsq}, the optimal number of latent factors $d$ for Foursquare's is 80. For higher values of latent factors, nDCG@5 drops. Also, nDCG@5 drops when selecting lower values for $d$. 
We observe a similar behavior on Gowalla's in \autoref{fig-dim-gwl} with the difference that the optimal number of latent factors is 90. For all other values of $d$, we observe a drop in the performance. 

\begin{figure}[t]
	\begin{subfigure}[b]{0.49\columnwidth}
        \includegraphics[height=0.15\textheight]{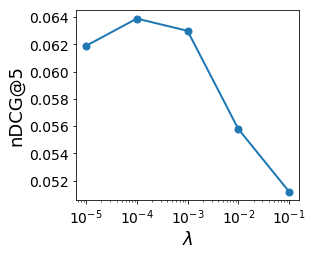}
        \caption{Foursquare's}
        \label{fig-lambda-fsq}
    \end{subfigure}
	\begin{subfigure}[b]{0.49\columnwidth}
        \includegraphics[height=0.15\textheight]{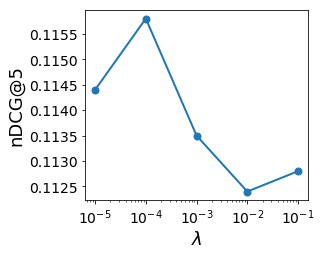}
        \caption{Gowalla's}
        \label{fig-lambda-gwl}
    \end{subfigure}
\caption{Effect of $\lambda$.} 
\label{fig-lambda}
\end{figure}

Furthermore, in order to study the effect of the regularizing control parameter ($\lambda$), we varied $\lambda$ while keeping $d$ and $\alpha$ fixed.
As shown in \autoref{fig-lambda-fsq}, the best $\lambda$ for Foursquare's is $\num{1e-4}$, while according to \autoref{fig-lambda-gwl}, for Gowalla's is $\num{1e-4}$. The performance of our model drops using different values of $\lambda$ for both datasets. While lower performance achieved with lower values of $\lambda$ indicate that the introduced regularizer is essential to avoid overfitting, higher values of $\lambda$ also hurt the performance. 

\begin{figure}[t]
	\begin{subfigure}[b]{0.49\columnwidth}
        \includegraphics[height=0.15\textheight]{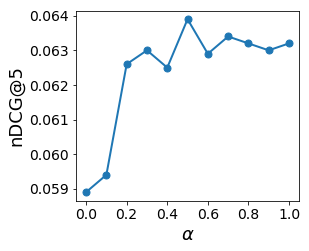}
        \caption{Foursquare's}
        \label{fig-alpha-fsq}
    \end{subfigure}
	\begin{subfigure}[b]{0.49\columnwidth}
        \includegraphics[height=0.15\textheight]{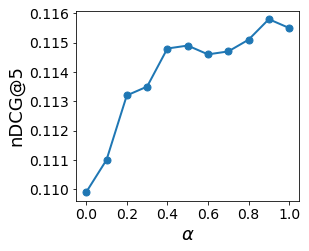}
        \caption{Gowalla's}
        \label{fig-alpha-gwl}
    \end{subfigure}
\caption{Effect of $\alpha$.} 
\label{fig-alpha}
\end{figure}

Next, in \autoref{fig-alpha} we study the effect of geographical influence weight $\alpha$ on the performance of our model and report nDCG@5 while keeping other parameters fixed.
As we can see in \autoref{fig-alpha-fsq}, the optimal value of $\alpha$ for Foursquare's is 0.5 and the performance drops for all other values of $\alpha$.
We observe a similar behavior on Gowalla's in \autoref{fig-alpha-gwl} where the best performance is achieved with $\alpha=0.9$. For all other values of $\alpha$, we see a drop in the performance.

\subsection{Model's Convergence}
\begin{figure}[t]
	\begin{subfigure}[b]{0.49\columnwidth}
        \includegraphics[height=0.15\textheight]{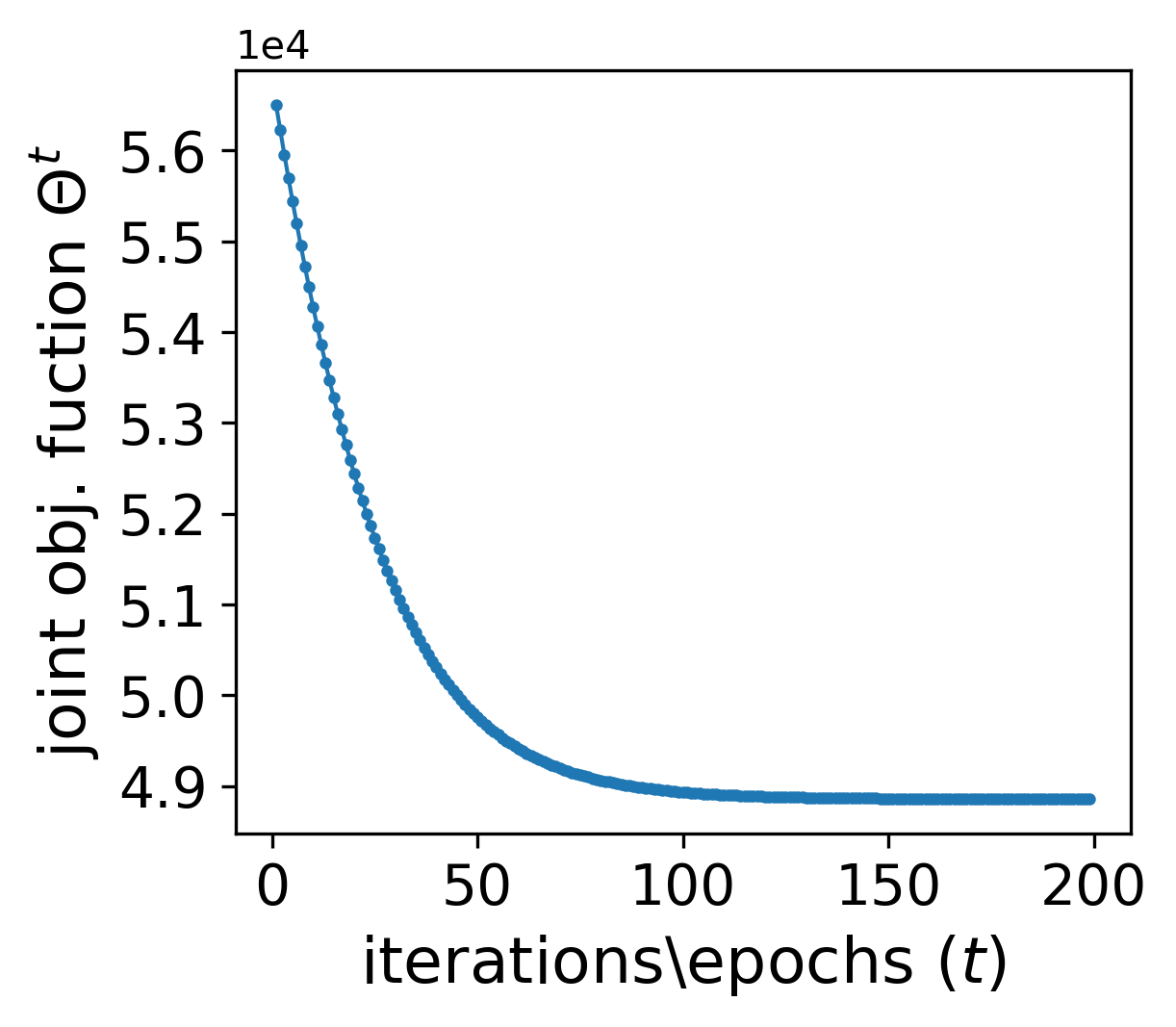}
        \caption{Foursquare's}
        \label{fig-convergence-fsq}
    \end{subfigure}
	\begin{subfigure}[b]{0.49\columnwidth}
        \includegraphics[height=0.15\textheight]{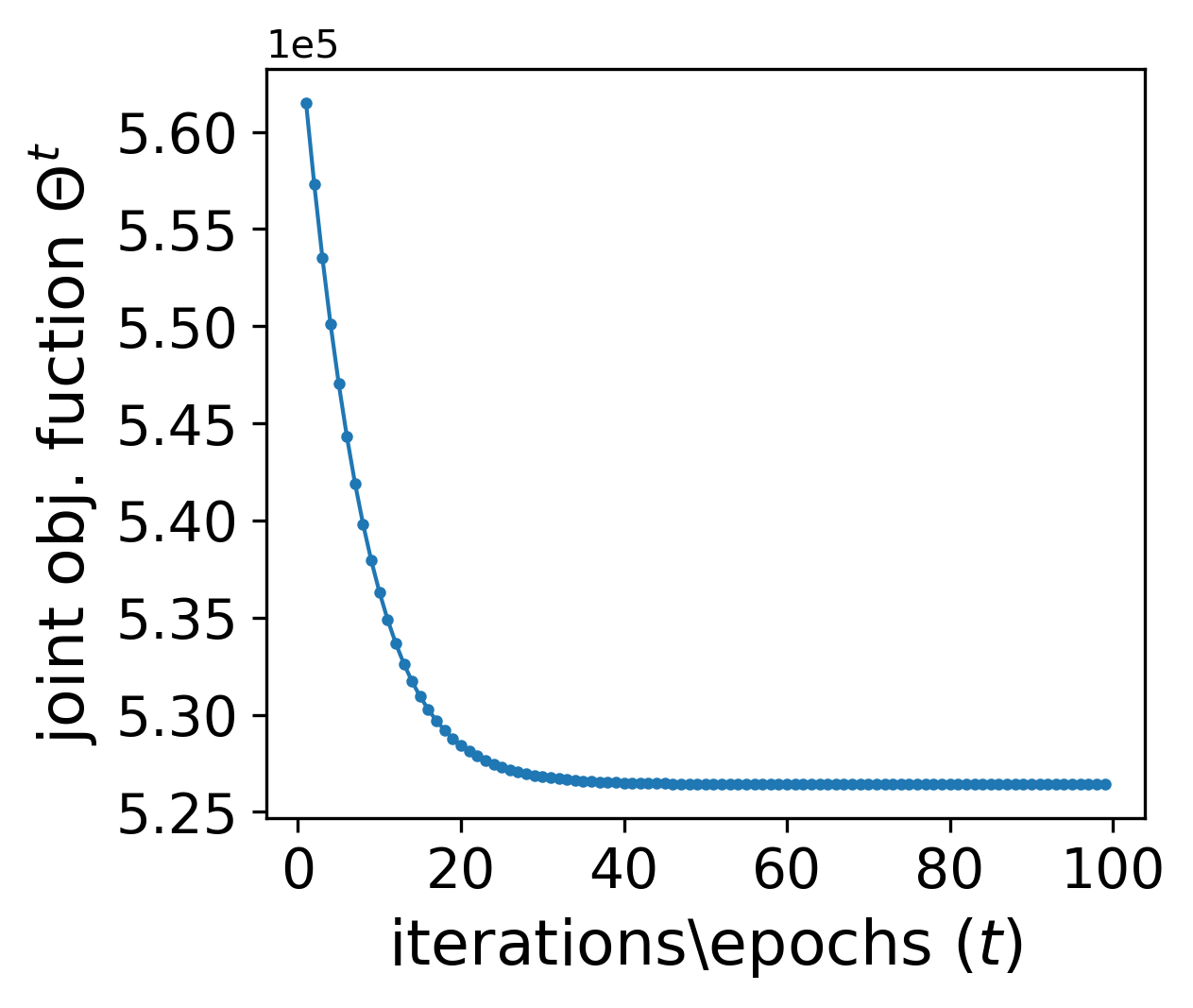}
        \caption{Gowalla's}
        \label{fig-convergence-gwl}
    \end{subfigure}
\caption{Convergence of the joint objective function.} 
\label{fig-convergence}
\end{figure}

Finally, Figure \ref{fig-convergence} plots the value of the joint objective function, $\Theta^t = R(U^t, V^t) + R_{\Pi}(U^t, V^t)$ (Eq. \eqref{eq-obj-r} and \eqref{eq-obj-rpi}) when training the model in $t$ iterations/epochs. We observe the value of $\Theta^t$ consistently decreases as the training epochs increase until the proposed \modelname model converges. The behavior of the proposed objective function is as expected, since it is the summation of a logistic loss and a quasi convex function. $R$, that is a logistic loss, is convex and monotonic decreasing. Also, $R_{\Pi}$ is strictly positive and monotonic decreasing. Hence, the summation of both loss functions converges as illustrated in Figure \ref{fig-convergence}.

\section{Conclusions and Future Work}
\label{sec-conclusion}

In this article, we presented an extensive data analysis of two POI recommendation datasets studying various attributes related to sparsity, time-sensitivity, and multiple check-ins. Based on the intuitions we got from data analysis, we proposed a two-phase CR model, called \modelname. In addition, we showed how to incorporate the geographical influence into the objective functions and proposed a time-sensitive regularizer to capture the long-term user behavior and POI popularity patterns.
The experimental results on the two benchmark datasets demonstrated that our proposed model outperforms other state-of-the-art methods. 
The results indicated that our model is able to address the data sparsity problem taking into account both visited and unvisited POIs in the training phase and their respective geographical distances. We also showed that the second phase is able to rank more-relevant POIs higher in the ranking, explaining the superiority of our two-phase model over the baselines as well as the first phase of the algorithm. This suggests that while single check-ins provide valuable information about the users' preferences, multiple check-ins give us a more clear picture of their behavior and habits. Therefore, in the first phase of our algorithm, a CR model that focuses on ranking visited venues higher than unvisited ones addresses the data sparsity by taking into account the unvisited venues in the training phase. In the second phase, a different CR approach is employed focusing on placing users' favorite POIs higher in the ranking. Throughout this process, we have regularized the learning procedure following the intuition that we had by analyzing time-sensitivity of users and POIs. Our aim was to penalize those POIs and users that have been more time-sensitive in the past. 

A wealth of information about users and POIs are available on various LBSNs. As future work, we plan to extend our two-phase model to generate POI recommendations considering users' behavior in different LBSNs~\cite{DBLP:conf/pkdd/RafailidisC16,DBLP:journals/tois/AliannejadiC18}. Previous works have shown that considering a cross-platform and multimodal behavioral analysis improves the performance of a model dramatically. Hence, we are very interested in investigating how we can extend our current work to consider multimodal information such as POI category, user reviews, and opening hours. 
Also, with the recent advances in applying deep neural models for POI recommendation~\cite{DBLP:conf/aaai/LiuWWT16,DBLP:conf/kdd/DuDTUGS16,DBLP:journals/tkde/YinWWCZ17,DBLP:conf/cikm/ManotumruksaMO17} and their power to capture complex structures of user-POI interactions, we plan to combine our joint learning approach with the existing deep recurrent neural models to explore its potential benefits to a deep neural recommender model. 
Furthermore, since our training strategy requires negative training examples, we considered all unseen POIs as negative samples which increases the complexity of the model. Inspired by the relevant studies \cite{DBLP:conf/cikm/ManotumruksaMO17a,DBLP:conf/aaai/ChengYKL12,DBLP:conf/sigir/YeYLL11}, we plan to explore various strategies for negative sampling and evaluate their effect on \modelname.

\section*{Acknowledgement}
This research was partially funded by the RelMobIR project of the Swiss National Science Foundation (SNSF).

\IEEEdisplaynontitleabstractindextext

\bibliographystyle{IEEEtran}
\bibliography{IEEEabrv,main}

\begin{thebibliography}{10}
\providecommand{\url}[1]{#1}
\csname url@samestyle\endcsname
\providecommand{\newblock}{\relax}
\providecommand{\bibinfo}[2]{#2}
\providecommand{\BIBentrySTDinterwordspacing}{\spaceskip=0pt\relax}
\providecommand{\BIBentryALTinterwordstretchfactor}{4}
\providecommand{\BIBentryALTinterwordspacing}{\spaceskip=\fontdimen2\font plus
\BIBentryALTinterwordstretchfactor\fontdimen3\font minus
  \fontdimen4\font\relax}
\providecommand{\BIBforeignlanguage}[2]{{%
\expandafter\ifx\csname l@#1\endcsname\relax
\typeout{** WARNING: IEEEtran.bst: No hyphenation pattern has been}%
\typeout{** loaded for the language `#1'. Using the pattern for}%
\typeout{** the default language instead.}%
\else
\language=\csname l@#1\endcsname
\fi
#2}}
\providecommand{\BIBdecl}{\relax}
\BIBdecl

\bibitem{DBLP:journals/umuai/ChenCW15}
L.~Chen, G.~Chen, and F.~Wang, ``Recommender systems based on user reviews: the
  state of the art,'' \emph{User Modeling and User-Adapted Interaction},
  vol.~25, no.~2, pp. 99--154, 2015.

\bibitem{DBLP:conf/cikm/YuanCS14}
Q.~Yuan, G.~Cong, and A.~Sun, ``Graph-based point-of-interest recommendation
  with geographical and temporal influences,'' in \emph{CIKM}, 2014, pp.
  659--668.

\bibitem{DBLP:journals/tois/AdomaviciusSST05}
G.~Adomavicius, R.~Sankaranarayanan, S.~Sen, and A.~Tuzhilin, ``Incorporating
  contextual information in recommender systems using a multidimensional
  approach,'' \emph{{ACM} Trans. Inf. Syst.}, vol.~23, no.~1, pp. 103--145,
  2005.

\bibitem{DBLP:conf/sac/AliannejadiMC17}
M.~Aliannejadi, I.~Mele, and F.~Crestani, ``Personalized ranking for
  context-aware venue suggestion,'' in \emph{SAC}, 2017, pp. 960--962.

\bibitem{DBLP:journals/tkde/AdomaviciusT05}
G.~Adomavicius and A.~Tuzhilin, ``Toward the next generation of recommender
  systems: {A} survey of the state-of-the-art and possible extensions,''
  \emph{{IEEE} Trans. Knowl. Data Eng.}, vol.~17, no.~6, pp. 734--749, 2005.

\bibitem{DBLP:journals/geoinformatica/0003ZWM15}
J.~Bao, Y.~Zheng, D.~Wilkie, and M.~F. Mokbel, ``Recommendations in
  location-based social networks: a survey,'' \emph{GeoInformatica}, vol.~19,
  no.~3, pp. 525--565, 2015.

\bibitem{DBLP:conf/www/ZhengZXY10}
V.~W. Zheng, Y.~Zheng, X.~Xie, and Q.~Yang, ``Collaborative location and
  activity recommendations with {GPS} history data,'' in \emph{WWW}, 2010.

\bibitem{DBLP:conf/icde/LevandoskiSEM12}
J.~J. Levandoski, M.~Sarwat, A.~Eldawy, and M.~F. Mokbel, ``{LARS:} {A}
  location-aware recommender system,'' in \emph{ICDE}, 2012, pp. 450--461.

\bibitem{DBLP:conf/cikm/FerenceYL13}
G.~Ference, M.~Ye, and W.~Lee, ``Location recommendation for out-of-town users
  in location-based social networks,'' in \emph{CIKM}, 2013, pp. 721--726.

\bibitem{DBLP:conf/recsys/GriesnerAN15}
J.~Griesner, T.~Abdessalem, and H.~Naacke, ``{POI} recommendation: Towards
  fused matrix factorization with geographical and temporal influences,'' in
  \emph{RecSys}, 2015, pp. 301--304.

\bibitem{DBLP:conf/sigir/LiCLPK15}
X.~Li, G.~Cong, X.~Li, T.~N. Pham, and S.~Krishnaswamy, ``Rank-geofm: {A}
  ranking based geographical factorization method for point of interest
  recommendation,'' in \emph{SIGIR}, 2015, pp. 433--442.

\bibitem{DBLP:conf/kdd/YeSLYJ11}
M.~Ye, D.~Shou, W.~Lee, P.~Yin, and K.~Janowicz, ``On the semantic annotation
  of places in location-based social networks,'' in \emph{KDD}, 2011, pp.
  520--528.

\bibitem{DBLP:conf/airs/AliannejadiMC16}
M.~Aliannejadi, I.~Mele, and F.~Crestani, ``User model enrichment for venue
  recommendation,'' in \emph{AIRS}, 2016, pp. 212--223.

\bibitem{DBLP:conf/kdd/LianZXSCR14}
D.~Lian, C.~Zhao, X.~Xie, G.~Sun, E.~Chen, and Y.~Rui, ``Geomf: joint
  geographical modeling and matrix factorization for point-of-interest
  recommendation,'' in \emph{KDD}, 2014, pp. 831--840.

\bibitem{DBLP:conf/nips/SalakhutdinovM07}
R.~Salakhutdinov and A.~Mnih, ``Probabilistic matrix factorization,'' in
  \emph{NIPS}, 2007, pp. 1257--1264.

\bibitem{DBLP:conf/www/Christakopoulou15}
K.~Christakopoulou and A.~Banerjee, ``Collaborative ranking with a push at the
  top,'' in \emph{WWW}, 2015, pp. 205--215.

\bibitem{DBLP:conf/wsdm/BalakrishnanC12}
S.~Balakrishnan and S.~Chopra, ``Collaborative ranking,'' in \emph{WSDM}, 2012,
  pp. 143--152.

\bibitem{DBLP:conf/nips/WeimerKLS07}
M.~Weimer, A.~Karatzoglou, Q.~V. Le, and A.~J. Smola, ``{COFI} {RANK} - maximum
  margin matrix factorization for collaborative ranking,'' in \emph{NIPS},
  2007, pp. 1593--1600.

\bibitem{DBLP:conf/uai/RendleFGS09}
S.~Rendle, C.~Freudenthaler, Z.~Gantner, and L.~Schmidt{-}Thieme, ``{BPR:}
  bayesian personalized ranking from implicit feedback,'' in \emph{UAI}, 2009,
  pp. 452--461.

\bibitem{DBLP:conf/sigir/YuanCMSM13}
Q.~Yuan, G.~Cong, Z.~Ma, A.~Sun, and N.~Magnenat{-}Thalmann, ``Time-aware
  point-of-interest recommendation,'' in \emph{SIGIR}, 2013, pp. 363--372.

\bibitem{DBLP:conf/recsys/GaoTHL13}
H.~Gao, J.~Tang, X.~Hu, and H.~Liu, ``Exploring temporal effects for location
  recommendation on location-based social networks,'' in \emph{RecSys}, 2013,
  pp. 93--100.

\bibitem{DBLP:conf/kdd/LiuLLQX16}
Y.~Liu, C.~Liu, B.~Liu, M.~Qu, and H.~Xiong, ``Unified point-of-interest
  recommendation with temporal interval assessment,'' in \emph{KDD}, 2016, pp.
  1015--1024.

\bibitem{DBLP:conf/ijcai/FengLZCCY15}
S.~Feng, X.~Li, Y.~Zeng, G.~Cong, Y.~M. Chee, and Q.~Yuan, ``Personalized
  ranking metric embedding for next new {POI} recommendation,'' in
  \emph{IJCAI}, 2015, pp. 2069--2075.

\bibitem{DBLP:journals/jmlr/Rudin09}
C.~Rudin, ``The p-norm push: {A} simple convex ranking algorithm that
  concentrates at the top of the list,'' \emph{Journal of Machine Learning
  Research}, vol.~10, pp. 2233--2271, 2009.

\bibitem{DBLP:journals/cacm/GoldbergNOT92}
D.~Goldberg, D.~A. Nichols, B.~M. Oki, and D.~B. Terry, ``Using collaborative
  filtering to weave an information tapestry,'' \emph{Commun. {ACM}}, vol.~35,
  no.~12, pp. 61--70, 1992.

\bibitem{DBLP:journals/tois/YinCSHC14}
H.~Yin, B.~Cui, Y.~Sun, Z.~Hu, and L.~Chen, ``{LCARS:} {A} spatial item
  recommender system,'' \emph{{ACM} Trans. Inf. Syst.}, vol.~32, no.~3, pp.
  11:1--11:37, 2014.

\bibitem{DBLP:conf/www/SarwarKKR01}
B.~M. Sarwar, G.~Karypis, J.~A. Konstan, and J.~Riedl, ``Item-based
  collaborative filtering recommendation algorithms,'' in \emph{WWW}, 2001, pp.
  285--295.

\bibitem{koren2008factorization}
Y.~Koren, ``Factorization meets the neighborhood: a multifaceted collaborative
  filtering model,'' in \emph{CIKM}, 2008, pp. 426--434.

\bibitem{DBLP:conf/sigir/YeYLL11}
M.~Ye, P.~Yin, W.~Lee, and D.~L. Lee, ``Exploiting geographical influence for
  collaborative point-of-interest recommendation,'' in \emph{SIGIR}, 2011, pp.
  325--334.

\bibitem{DBLP:journals/tist/ChengYKL16}
C.~Cheng, H.~Yang, I.~King, and M.~R. Lyu, ``A unified point-of-interest
  recommendation framework in location-based social networks,'' \emph{{ACM}
  Trans. Intell. Sys. and Tech.}, vol.~8, no.~1, pp. 10:1--10:21, 2016.

\bibitem{DBLP:journals/tois/ZhangLW16}
C.~Zhang, H.~Liang, and K.~Wang, ``Trip recommendation meets real-world
  constraints: {POI} availability, diversity, and traveling time uncertainty,''
  \emph{{ACM} Trans. Inf. Syst.}, vol.~35, no.~1, pp. 5:1--5:28, 2016.

\bibitem{DBLP:conf/ecir/Aliannejadi17}
M.~Aliannejadi, D.~Rafailidis, and F.~Crestani, ``Personalized keyword boosting
  for venue suggestion based on multiple {LBSNs},'' in \emph{ECIR}, 2017, pp.
  291--303.

\bibitem{alianSigir17}
M.~Aliannejadi and F.~Crestani, ``Venue appropriateness prediction for
  personalized context-aware venue suggestion,'' in \emph{SIGIR}, 2017, pp.
  1177--1180.

\bibitem{DBLP:journals/tist/ZhangDCLZ13}
W.~Zhang, G.~Ding, L.~Chen, C.~Li, and C.~Zhang, ``Generating virtual ratings
  from chinese reviews to augment online recommendations,'' \emph{{ACM} Trans.
  Intell. Sys. and Tech.}, vol.~4, no.~1, pp. 9:1--9:17, 2013.

\bibitem{DBLP:journals/tist/CuiSNHM17}
C.~Cui, J.~Shen, L.~Nie, R.~Hong, and J.~Ma, ``Augmented collaborative
  filtering for sparseness reduction in personalized {POI} recommendation,''
  \emph{{ACM} Trans. Intell. Sys. and Tech.}, vol.~8, no.~5, pp. 71:1--71:23,
  2017.

\bibitem{DBLP:conf/ictai/YuanJGCYA16}
F.~Yuan, J.~M. Jose, G.~Guo, L.~Chen, H.~Yu, and R.~S. Alkhawaldeh, ``Joint
  geo-spatial preference and pairwise ranking for point-of-interest
  recommendation,'' in \emph{ICTAI}, 2016, pp. 46--53.

\bibitem{DBLP:journals/ftir/Liu09}
T.~Liu, ``Learning to rank for information retrieval,'' \emph{Foundations and
  Trends in Information Retrieval}, vol.~3, no.~3, pp. 225--331, 2009.

\bibitem{DBLP:conf/nips/CrammerS01}
K.~Crammer and Y.~Singer, ``Pranking with ranking,'' in \emph{NIPS}, 2001, pp.
  641--647.

\bibitem{rankNet}
C.~J.~C. Burges, T.~Shaked, E.~Renshaw, A.~Lazier, M.~Deeds, N.~Hamilton, and
  G.~N. Hullender, ``Learning to rank using gradient descent,'' in \emph{ICML},
  2005, pp. 89--96.

\bibitem{listNet}
Z.~Cao, T.~Qin, T.~Liu, M.~Tsai, and H.~Li, ``Learning to rank: from pairwise
  approach to listwise approach,'' in \emph{ICML}, 2007, pp. 129--136.

\bibitem{DBLP:conf/cikm/ShiKBLH13}
Y.~Shi, A.~Karatzoglou, L.~Baltrunas, M.~Larson, and A.~Hanjalic, ``Gapfm:
  optimal top-n recommendations for graded relevance domains,'' in \emph{CIKM},
  2013, pp. 2261--2266.

\bibitem{DBLP:conf/sigir/RafailidisC16a}
D.~Rafailidis and F.~Crestani, ``Collaborative ranking with social
  relationships for top-n recommendations,'' in \emph{SIGIR}, 2016, pp.
  785--788.

\bibitem{DBLP:conf/cikm/RafailidisC16}
------, ``Joint collaborative ranking with social relationships in top-n
  recommendation,'' in \emph{CIKM}, 2016, pp. 1393--1402.

\bibitem{DBLP:conf/recsys/RafailidisC17}
------, ``Learning to rank with trust and distrust in recommender systems,'' in
  \emph{RecSys}, 2017, pp. 5--13.

\bibitem{DBLP:conf/cikm/DingL05}
Y.~Ding and X.~Li, ``Time weight collaborative filtering,'' in \emph{CIKM},
  2005, pp. 485--492.

\bibitem{DBLP:conf/icdm/YaoFLLX16}
Z.~Yao, Y.~Fu, B.~Liu, Y.~Liu, and H.~Xiong, ``{POI} recommendation: {A}
  temporal matching between {POI} popularity and user regularity,'' in
  \emph{ICDM}, 2016, pp. 549--558.

\bibitem{DBLP:journals/tois/LiJHL17}
X.~Li, M.~Jiang, H.~Hong, and L.~Liao, ``A time-aware personalized
  point-of-interest recommendation via high-order tensor factorization,''
  \emph{{ACM} Trans. Inf. Syst.}, vol.~35, no.~4, pp. 31:1--31:23, 2017.

\bibitem{DBLP:journals/tois/YinCZWHS16}
H.~Yin, B.~Cui, X.~Zhou, W.~Wang, Z.~Huang, and S.~W. Sadiq, ``Joint modeling
  of user check-in behaviors for real-time point-of-interest recommendation,''
  \emph{{ACM} Trans. Inf. Syst.}, vol.~35, no.~2, pp. 11:1--11:44, 2016.

\bibitem{DBLP:conf/sigmod/YinCCHH14}
H.~Yin, B.~Cui, L.~Chen, Z.~Hu, and Z.~Huang, ``A temporal context-aware model
  for user behavior modeling in social media systems,'' in \emph{SIGMOD}, 2014,
  pp. 1543--1554.

\bibitem{DBLP:journals/tois/YinCCHZ15}
H.~Yin, B.~Cui, L.~Chen, Z.~Hu, and X.~Zhou, ``Dynamic user modeling in social
  media systems,'' \emph{{ACM} Trans. Inf. Syst.}, vol.~33, no.~3, pp.
  10:1--10:44, 2015.

\bibitem{DBLP:conf/ijcai/ChengYLK13}
C.~Cheng, H.~Yang, M.~R. Lyu, and I.~King, ``Where you like to go next:
  Successive point-of-interest recommendation,'' in \emph{IJCAI}, 2013, pp.
  2605--2611.

\bibitem{DBLP:conf/iir/AliannejadiC18}
M.~Aliannejadi and F.~Crestani, ``A collaborative ranking model with contextual
  similarities for venue suggestion,'' in \emph{{IIR}}, 2018.

\bibitem{DBLP:conf/ictir/AliannejadiRC18}
M.~Aliannejadi, D.~Rafailidis, and F.~Crestani, ``A collaborative ranking model
  with multiple location-based similarities for venue suggestion,'' in
  \emph{{ICTIR}}, 2018, pp. 19--26.

\bibitem{DBLP:conf/ecir/KoolenBBK15}
M.~Koolen, T.~Bogers, A.~van~den Bosch, and J.~Kamps, ``Looking for books in
  social media: An analysis of complex search requests,'' in \emph{ECIR}, 2015,
  pp. 184--196.

\bibitem{DBLP:conf/www/ZhaoZKL17}
S.~Zhao, T.~Zhao, I.~King, and M.~R. Lyu, ``Geo-teaser: Geo-temporal sequential
  embedding rank for point-of-interest recommendation,'' in \emph{WWW}, 2017,
  pp. 153--162.

\bibitem{DBLP:conf/icdm/HuKV08}
Y.~Hu, Y.~Koren, and C.~Volinsky, ``Collaborative filtering for implicit
  feedback datasets,'' in \emph{ICDM}, 2008, pp. 263--272.

\bibitem{DBLP:conf/cikm/LiuWSM14}
Y.~Liu, W.~Wei, A.~Sun, and C.~Miao, ``Exploiting geographical neighborhood
  characteristics for location recommendation,'' in \emph{CIKM}, 2014, pp.
  739--748.

\bibitem{DBLP:conf/icdm/PanZCLLSY08}
R.~Pan, Y.~Zhou, B.~Cao, N.~N. Liu, R.~M. Lukose, M.~Scholz, and Q.~Yang,
  ``One-class collaborative filtering,'' in \emph{ICDM}, 2008, pp. 502--511.

\bibitem{DBLP:conf/pkdd/RafailidisC16}
D.~Rafailidis and F.~Crestani, ``Top-n recommendation via joint cross-domain
  user clustering and similarity learning,'' in \emph{{ECML/PKDD}}, 2016, pp.
  426--441.

\bibitem{DBLP:journals/tois/AliannejadiC18}
M.~Aliannejadi and F.~Crestani, ``Personalized context-aware point of interest
  recommendation,'' \emph{{ACM} Trans. Inf. Syst.}, vol.~36, no.~4, pp.
  45:1--45:28, 2018.

\bibitem{DBLP:conf/aaai/LiuWWT16}
Q.~Liu, S.~Wu, L.~Wang, and T.~Tan, ``Predicting the next location: {A}
  recurrent model with spatial and temporal contexts,'' in \emph{AAAI}, 2016,
  pp. 194--200.

\bibitem{DBLP:conf/kdd/DuDTUGS16}
N.~Du, H.~Dai, R.~Trivedi, U.~Upadhyay, M.~Gomez{-}Rodriguez, and L.~Song,
  ``Recurrent marked temporal point processes: Embedding event history to
  vector,'' in \emph{KDD}, 2016, pp. 1555--1564.

\bibitem{DBLP:journals/tkde/YinWWCZ17}
H.~Yin, W.~Wang, H.~Wang, L.~Chen, and X.~Zhou, ``Spatial-aware hierarchical
  collaborative deep learning for {POI} recommendation,'' \emph{{IEEE} Trans.
  Knowl. Data Eng.}, vol.~29, no.~11, pp. 2537--2551, 2017.

\bibitem{DBLP:conf/cikm/ManotumruksaMO17}
J.~Manotumruksa, C.~Macdonald, and I.~Ounis, ``A deep recurrent collaborative
  filtering framework for venue recommendation,'' in \emph{CIKM}, 2017, pp.
  1429--1438.

\bibitem{DBLP:conf/cikm/ManotumruksaMO17a}
------, ``A personalised ranking framework with multiple sampling criteria for
  venue recommendation,'' in \emph{{CIKM}}, 2017, pp. 1469--1478.

\bibitem{DBLP:conf/aaai/ChengYKL12}
C.~Cheng, H.~Yang, I.~King, and M.~R. Lyu, ``Fused matrix factorization with
  geographical and social influence in location-based social networks,'' in
  \emph{{AAAI}}, 2012, pp. 17--23.

\end{thebibliography}
\vskip -3\baselineskip plus -1fil
\begin{IEEEbiography}[{\includegraphics[width=1in,height=1.10in,clip,keepaspectratio]{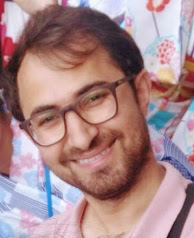}}]{Mohammad Aliannejadi} is a Ph.D. candidate at the Universit{\`a} della Svizzera italiana (USI) in Lugano, Switzerland, working on Mobile Information Retrieval as well as POI Recommendation. His main emphasis is on studying cross-app mobile search. Before that, he was an M.Sc. student at Amirkabir University of Technology (Tehran Polytechnic) in Tehran, Iran, where he worked on Spoken Language Understanding.
\end{IEEEbiography}
\vskip -3\baselineskip plus -1fil
\begin{IEEEbiography}[{\includegraphics[width=1in,height=1.10in,clip,keepaspectratio]{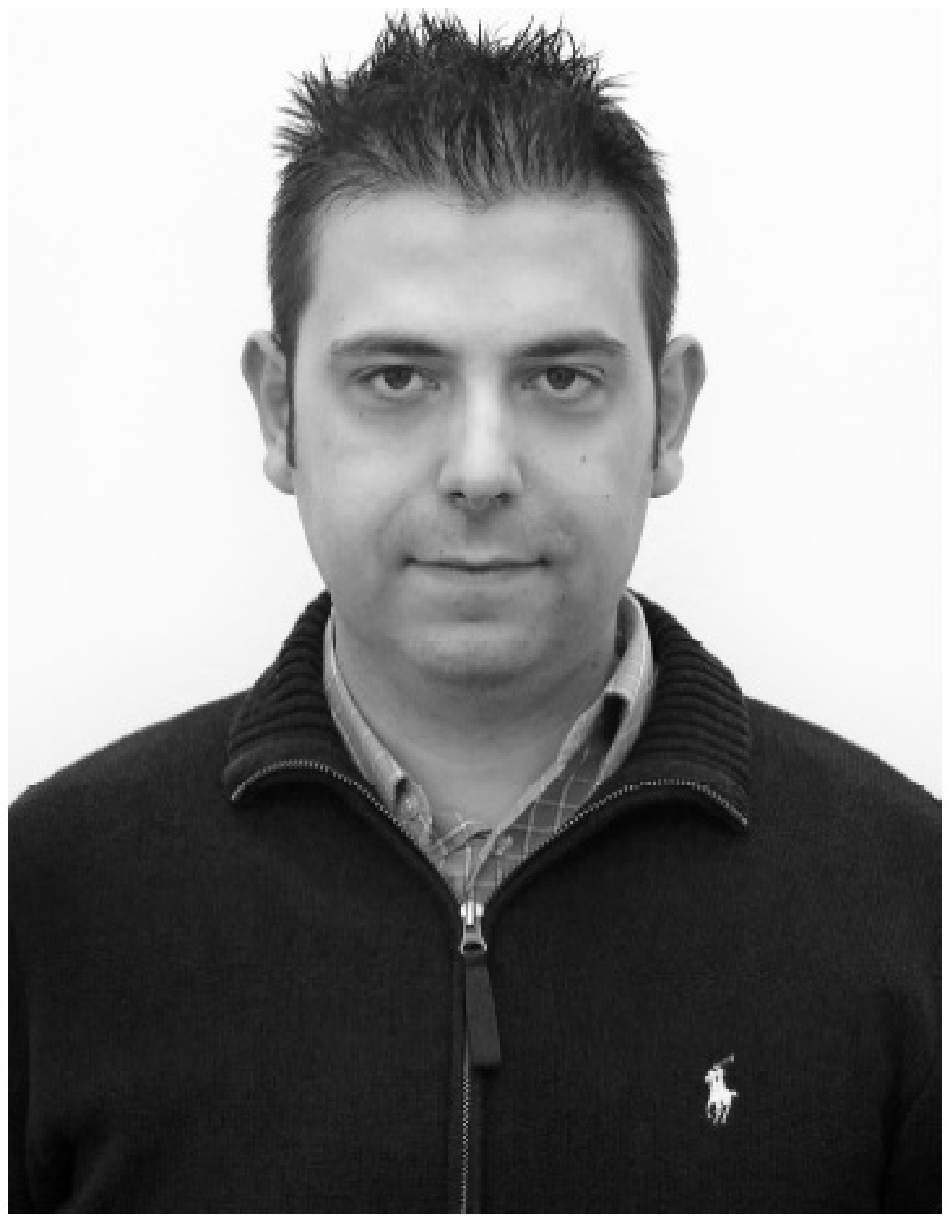}}]{Dimitrios Rafailidis} received the B.Sc., M.Sc., and Ph.D. degrees from the Department of Informatics of the Aristotle University of Thessaloniki, Greece, in 2005, 2007, and 2011, respectively. 
He is an Assistant Professor at Maastricht Universiy. His research interests include Machine Learning, Information Retrieval, Recommender Systems and Social Media Mining.
\end{IEEEbiography}
\vskip -3\baselineskip plus -1fil
\begin{IEEEbiography}[{\includegraphics[width=1in,height=1.10in,clip,keepaspectratio]{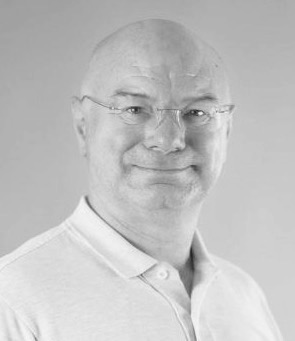}}]{Fabio Crestani} is a Full Professor at the Faculty of Informatics of the Universit{\`a} della Svizzera italiana (USI) in Lugano, Switzerland, since 2007.  Previously he was a Professor at the University of Strathclyde in Glasgow, United Kingdom. Prof Crestani is an internationally recognized researcher in Information Retrieval, Text Mining and Digital Libraries. In these areas he has published over two hundred refereed papers on both theoretical and experimental topics. 
He was the Editor-in-Chief of Information Processing and Management until 2015, has served in the organizing and program committee of several conferences and in the editorial board of several journals. 
\end{IEEEbiography}

\end{document}